\documentclass[useAMS,usenatbib]{mn2e}
\usepackage{graphicx}
\usepackage{aas_macros}

\title[Statistical uncertainties in nebular abundances]{Understanding and reducing statistical uncertainties in nebular abundance determinations} %have to be the same?
\author[R. Wesson et al.]{R. Wesson$^{1,2}$, D.J. Stock$^{1,3}$ \& P. Scicluna$^{1,4}$\\
$^1$Department of Physics and Astronomy, University College London, Gower Street, London WC1E 6BT, UK\\
$^2$European Southern Observatory, Alonso de C\'ordova 3107, Casilla 19001, Santiago, Chile\\
$^3$Department of Physics and Astronomy, University of Western Ontario, London, Ontario, Canada, N6K 3K7\\
$^4$European Southern Observatory, Karl-Schwarzschildstr. 2, 85748 Garching, Germany\\ 
}

\begin{document}

\date{}

\pagerange{\pageref{firstpage}--\pageref{lastpage}} \pubyear{2002}

\maketitle

\label{firstpage}

\begin{abstract}
Whenever observations are compared to theories, an estimate of the uncertainties associated with the observations is vital if the comparison is to be meaningful.  However, many or even most determinations of temperatures, densities and abundances in photoionized nebulae do not quote the associated uncertainty.  Those that do typically propagate the uncertainties using analytical techniques which rely on assumptions that generally do not hold.

Motivated by this issue, we have developed {\sc neat} (Nebular Empirical Analysis Tool), a new code for calculating chemical abundances in photoionized nebulae.  The code carries out a standard analysis of lists of emission lines using long-established techniques to estimate the amount of interstellar extinction, calculate representative temperatures and densities, compute ionic abundances from both collisionally excited lines and recombination lines, and finally to estimate total elemental abundances using an ionization correction scheme.  {\sc neat} uses a Monte Carlo technique to robustly propagate uncertainties from line flux measurements through to the derived abundances.

We show that for typical observational data, this approach is superior to analytic estimates of uncertainties.  {\sc neat} also accounts for the effect of upward biasing on measurements of lines with low signal to noise, allowing us to accurately quantify the effect of this bias on abundance determinations.  We find not only that the effect can result in significant over-estimates of heavy element abundances derived from weak lines, but that taking it into account reduces the uncertainty of these abundance determinations.  Finally, we investigate the effect of possible uncertainties in R, the ratio of selective to total extinction, on abundance determinations.  We find that the uncertainty due to this parameter is negligible compared to the statistical uncertainties due to typical line flux measurement uncertainties.

\end{abstract}


\section{Introduction}

Abundance determinations from photoionized nebulae play crucial roles in a variety of astrophysical contexts. Nebulae around evolved stars, e.g. Planetary Nebulae (PNe) or Wolf-Rayet (WR) ejecta nebulae provide constraints on theories of stellar nucleosynthesis and evolution (e.g. \citealt{2009ApJ...690.1130K};  \citealt{2011arXiv1110.1186M}; \citealt{1992A&A...264..105M}; \citealt{2011MNRAS.418.2532S}). Such data is invaluable as constraints on stellar yields; the \textit{inputs} of galactic chemical evolution models. Meanwhile, Galactic and extragalactic H~{\sc ii} regions provide insights into the current composition of the ISM and therefore are vital constraints for the \textit{output} of galactic chemical evolution models (e.g. \citealt{1997nceg.book.....P}; \citealt{2003ceg..book.....M}).

Studies of photoionized nebulae have a very long history, stretching back to the dawn of astrophysics (e.g. \citealt{1864RSPT..154..437H}), but some major questions remain unanswered.  One example is the sometimes sizeable discrepancy between abundances derived from collisionally excited lines (CELs) and those derived from recombination lines (RLs) (\citealt{2005MNRAS.362..424W}; \citealt{2006MNRAS.368.1959L}; \citealt{2007ApJ...670..457G}; \citealt{2008MNRAS.386...22T}).

To meaningfully assess whether results determined observationally are consistent or discrepant with model predictions, the uncertainties on both the observations and the model must be estimated.  In neither case is such an estimate straightforward.  The uncertainty on the observations is a combination of statistical uncertainties, relating ultimately to measurement uncertainties, and systematic uncertainties, which may arise from many sources including the instrumentation used to obtain the line flux measurements, the chosen parametrisation of the interstellar reddening law, and the methodological choices made in the analysis.  Sources of systematic uncertainty may be unknown and even unquantifiable.  Finally, the uncertainty attributed to models is perhaps the most difficult to quantify, being related ultimately to the confidence the modeller has that the model encapsulates the underlying physics without excessive simplification or unwarranted assumptions.

We concern ourselves in this paper with an improved understanding of statistical uncertainties and their effect on empirically determined nebular abundances.  We present a new code for calculating chemical abundances in photoionized nebulae, which can also robustly estimate statistical uncertainties using a Monte Carlo approach.  This method is inherently superior to analytic methods of uncertainty propagation, which rely on assumptions which are almost always violated by the measurement uncertainties inherent to actual astronomical observations.

We use our code to reanalyse several published line lists for which uncertainties are available, and we show firstly that analytic methods do not give a good representation of the true uncertainties on derived quantities.  Generally, uncertainties on derived abundances are better characterised by log-normal distributions than by normal distributions.  In some cases, bimodal probability distributions emerge.

Secondly, as the Monte Carlo approach can trivially handle any quantifiable distribution of line flux uncertainties, we have designed the code to account for the well known upward bias in measurements of weak lines, and associated log-normal distribution of measurement uncertainties (\citealt{1994A&A...287..676R}).  The bias is well known but its actual effect on abundance determinations has not previously been rigorously quantified.  We show, as expected, that abundances determined from weak lines are systematically overestimated.  Moreover, we show that in addition to removing this bias, accounting for the effect leads to a significant reduction in the uncertainty of abundance determinations from weak lines.

Finally, we investigate the effect of an assumed statistical uncertainty in R, the ratio of selective to total extinction.  We show that the extra uncertainty introduced into abundance determinations by taking this parameter to be 3.1 $\pm$ 0.15 instead of a fixed value of 3.1 is negligible compared to the uncertainties arising from line flux measurements, even when the extinction is quite large and the line fluxes are well measured.

The words ``uncertainty" and ``error" are often used synonymously.  In this article we maintain a distinction in meaning between the terms: ``uncertainty" refers to the limiting accuracy of the knowledge of a quantity, while ``error" refers to an actual mistake.

\section{Statistical uncertainties}

Uncertainties in observed quantities can be propagated into the uncertainty on derived parameters in a number of different ways, the two most common of which are the traditional analytical technique based on systems of partial derivatives and simplifying assumptions that allow one to apply Taylor expansions, and the `Monte Carlo' method which is a brute-force iterative method that exploits the wealth of computational power now readily available by building on knowledge of the uncertainty in the original observations.

The analytical approach is as follows: if one has measured a quantity $x$ with some uncertainty $\sigma x$, and wishes to calculate the uncertainty in a quantity $F$ given that $F = f(x)$, then the uncertainty on $F$ can be computed via the relation 

\begin{equation}
  \frac{\sigma F}{\sigma x} \simeq \frac{\partial f}{\partial x}
\end{equation}

However, in general one may not be able to compute this partial differential exactly, and in these cases, provided that

\begin{equation}
  \frac{\partial f}{\partial x}|_{x=x_1} \ll f(x_1)
\end{equation}

it is possible to use a first order Taylor expansion to approximate this derivative.  If one had a third value, $H = h(f, g)$ where $f$ and $g$ are both functions of other variables and are statistically independent of one another, it would be necessary to find the total derivative $dH$ such that

\begin{equation}
dh^2 = \left(\frac{\partial h}{\partial f}\right)^2df^2 + \left(\frac{\partial h}{\partial g}\right)^2dg^2
\end{equation}

and it thus follows that

\begin{equation}
\sigma^2_H = \left(\frac{\partial h}{\partial f}\sigma f\right)^2 + \left(\frac{\partial h}{\partial g}\sigma g\right)^2
\end{equation}

This expression can be generalised to any number of variables, and gives rise to the usual quadrature formulae for many simple functions of $x$.  We highlight three key aspects of this approach:

\begin{enumerate}
  \item Given the number of formulae through which the original line flux data must be put before an abundance can be determined, and the wide variety of functions applied, the equations necessary to propagate uncertainties in this way can become extremely complex;
  \item the approach implicitly assumes that all the input and output uncertainties at each step are normally distributed;
  \item the Taylor expansion requires that the uncertainties be small relative to the quantities.
\end{enumerate}

The first point is a matter of convenience but none the less one which discourages many authors from even attempting to propagate uncertainties all the way into the final quantities.  The second and third points are clearly violated in many or most real astrophysical observations, by virtue of which the uncertainties estimated using analytical techniques are liable not to reflect the true uncertainties.

As an example of the cumbersome nature of the analytic approach, we consider the estimation of the uncertainty on c(H$\beta$).  As described in Section~\ref{extinctionsection}, our code calculates c(H$\beta$) from the flux-weighted average of values derived from H$\alpha$, H$\gamma$ and H$\delta$.  Using the analytic approach described above, one finds the uncertainty on the unweighted c(H$\beta$) as follows.  Firstly,

\begin{equation}
c(H\beta) = \frac{I(H\alpha)c(H\beta)_\alpha + I(H\gamma)c(H\beta)_\gamma + I(H\delta)c(H\beta)_\delta}{I(H\alpha) + I(H\gamma) + I(H\delta)}
\end{equation}

where

\begin{equation}
c(H\beta)_\alpha = \frac{log\left(\frac{I(H\alpha)}{I_0(H\alpha)}\right)}{f(H\alpha)}
\end{equation}

and similarly for c(H$\beta)_\gamma$ and c(H$\beta)_\delta$.  The analytically estimated uncertainty on c(H$\beta$) derived using this method is

\begin{eqnarray}\nonumber
\frac{\Delta c(H\beta)}{c(H\beta} &= \bigg(\frac{(\Delta c(H\beta)_\alpha)^2 + (\Delta c(H\beta)_\gamma)^2+(\Delta c(H\beta)_\delta)^2}{(c(H\beta)_\alpha + c(H\beta)_\gamma + c(H\beta)_\delta)^2}\\
&+ \frac{(\Delta I(H\alpha)^2 + \Delta I(H\gamma)^2 + \Delta I(H\delta))^2}{(I(H\alpha) + I(H\gamma) + I(H\delta))^2}\bigg)^{0.5}
\end{eqnarray}

where

\begin{eqnarray}
\Delta c(H\beta)_\alpha &= c(H\beta)_\alpha \times \bigg(\left(\frac{\Delta I(H\alpha)}{I(H\alpha)}\right)^2 \\
&+ \left(\frac{0.434 \times (\Delta\frac{I(H\alpha)}{I_0(H\alpha)})}{\frac{I(H\alpha)}{I_0(H\alpha)}}\right)^2\bigg)^{0.5},
\end{eqnarray}

and similarly for $\Delta c(H\beta)_\gamma$ and $\Delta c(H\beta)_\delta$.  The mere derivation of such a formula is tiresome and complicated, and it would be exceedingly easy for errors to be introduced.  Even this is somewhat simplified: this equation assumes zero uncertainty in the values of f(H$\alpha, \gamma, \delta$) and I$_0(H\alpha, \gamma, \delta)$.  The uncertainty estimated for c(H$\beta$) then has to be propagated into the complex functions of line ratios which are used to give temperatures and densities, and then into the calculations of ionic abundances.

The Monte Carlo method, on the other hand, avoids all of the pitfalls of the analytical method.  It exploits the fact that if an observation of a quantity is drawn from a distribution $X$, with mean $x$ and variance $\sigma_x$ , and if one knows (or can make sensible assumptions about) the shape of this distribution, then a random-number generator can be used to repeatedly draw values from X, creating a random sample from it. Using the above example of $F = f (x)$, if one wanted to examine the uncertainty in $F$, the operation f(x) could be performed upon every value in the aforementioned sample to produce a sample of the distribution from which $F$ is drawn, which can then be parametrised to estimate the type, mean and standard deviation of this distribution. This process can be repeated ad infinitum for any number or combination of functions of $F$, $x$, or any other variable derived in the same way to propagate the uncertainties on the quantities, irrespective of the size of the uncertainties and any statistical interdependence of the variables.  This approach thus has the following advantages over the analytical approach:

\begin{enumerate}
  \item It is inherently very simple;
  \item Any distribution of uncertainties can be propagated at any stage;
  \item it does not require relative uncertainties to be small
\end{enumerate}

The Monte Carlo approach is thus inherently robust when applied to real astrophysical data, in a way that the analytical approach is not.  The only limitation is then the time taken to run the calculation enough times to sample well the statistics of the output distributions.

\section{{\sc neat}: Nebular empirical abundance tool}

We have developed a new code, {\sc neat} (Nebular Empirical Analysis Tool), to quickly carry out a thorough analysis of emission line spectra, and to propagate statistical uncertainties using the Monte Carlo technique described above.  In this section we describe the code and how it works.

\subsection{Input}

{\sc neat} incorporates elements of several previous codes, most importantly the {\sc equib} code, also developed at UCL, for solving the equations of statistical equilibrium in multi-level atoms.  All the source code, documentation, atomic data and example line lists are freely available at \texttt{http://www.sc.eso.org/\~{}rwesson/codes/neat/}.  The code is designed to be as simple as possible from the user's perspective, and our aim is that the user should be able to simply pass the code a line list and get out abundances determined by an objective methodology in which the user need not make any choices.  The code has no external dependencies and should compile without problems on any Unix-based system.  We have also verified that it compiles and runs on Windows systems, should the user be restricted to such an OS.

\subsubsection{Atomic data}

Hydrogen recombination data from \citet{1995MNRAS.272...41S} and helium recombination data from \citet{1996MNRAS.278..683S} are used throughout.  Helium abundances are corrected for collisional effects using the formulae provided by \citet{1995ApJ...442..714K}.

All atomic data for heavier elements is stored externally in plain text files, so that the user can easily change the atomic data being used without needing to edit or recompile the code.  We provide three sets of atomic data for collisionally excited lines with the code - a compilation from a variety of sources compiled on an ad hoc basis, CHIANTI 5.2 \citep{2006ApJS..162..261L}, and CHIANTI 6.0 \citep{2009A&A...498..915D}.  In all the analyses presented in this paper, we used atomic data from CHIANTI 5.2, with the exception of data for O$^{+}$ for which a documented error exists in the CHIANTI 5.2 data \citep{2009MNRAS.397..903K}, and S$^{2+}$, which we believe is affected by a similar error.  For O$^{+}$ we used transition probabilities from \citet{1982MNRAS.198..111Z} and collision strengths from \citet{1976MNRAS.177...31P}; for S$^{2+}$ we used transition probabilities from \citet{1982MNRAS.199.1025M} and collision strengths from \citet{1983IAUS..103..143M}.  For recombination lines, we use data from the sources given in Table~\ref{RL_atomic_data}.

\begin{table}
\begin{tabular}{ll}
\hline
Ion & Recombination data source \\
\hline
C$^{2+}$ & \citet{2000AAS..142...85D} \\
C$^{3+}$ & \citet{1991AA...251..680P} \\
N$^{2+}$ & \citet{1990ApJS...73..513E} \\
N$^{3+}$ & \citet{1991AA...251..680P} \\
O$^{2+}$ (3s--3p) & \citet{1994AA...282..999S} \\
O$^{2+}$ (3p--3d, 3d--4f) & \citet{1995MNRAS.272..369L} \\
Ne$^{2+}$ (3s--3p)& \citet{1998AAS..133..257K} \\
Ne$^{2+}$ (3d-4f) & Storey (unpublished) \\
\hline
\end{tabular}
\caption{Atomic data used in {\sc neat} for recombination lines}
\label{RL_atomic_data}
\end{table}

\subsubsection{Line list format}

{\sc neat} requires as input a plain text list of rest wavelengths, line fluxes, and uncertainties.  {\sc neat} currently recognises 738 lines, 81 of which are collisionally excited lines and 657 of which are recombination lines.  The code assigns line IDs based on an exact match to its reference list of rest wavelengths.  Different sources of atomic data often quote slightly different rest wavelengths for a given line; we include with {\sc NEAT} a separate code to read in line lists and reassign rest wavelengths to the values which {\sc NEAT} recognises.  Lines which are not recognised are ignored in the analysis.

Some cases of line blends can be accounted for; for example, the [O~{\sc ii}] lines at 3726 and 3729{\AA} may be blended in low resolution spectra.  In this case, a rest wavelength of 3727.00 can be given, and the code will properly treat the combined flux.  Blends of recombination lines cannot currently be accounted for but we plan to develop means of doing so in future versions of {\sc neat}.

For a very small number of recombination lines, close coincidences in rest wavelengths may lead to misidentifications.  For example, three weak O~{\sc ii} 3d--4f transitions from multiplets V63c, V78b and V63c have rest wavelengths of 4315.39, 4315.39 and 4315.40 respectively.  These three lines will almost certainy be detected as a blend if detected at all.  At the moment, {\sc neat} would attribute all flux at a wavelength of 4315.39 to each of the three lines.  The effect on final abundances should this occur is likely to be very small, as only very weak RLs are affected.  Again, we plan to improve the sophistication of {\sc neat}'s approach in future versions.

The user can select the number of iterations of the code to run.  If the number of iterations is one, the code performs a standard empirical analysis on the line list, as described below, and does not calculate any uncertainties.  If the number of iterations is more than one, then the code randomises the line list before each iteration.  The standard analysis is then carried out on the randomised line list.

\subsubsection{Line flux randomisation}
\label{randomising}

The code randomises the line fluxes by assuming that they come from one of three distributions, depending on the measured signal to noise.  The three cases are:

{\bf SNR$>$6.0: } the line flux is assumed to be drawn from a Gaussian distribution.  In this case, if the given line flux measurement is $F$ and the given measurement uncertainty is $\sigma$, then in each iteration of the code the line flux is calculated using

\begin{equation}
F_i = F + (R\times\sigma)
\end{equation}

where R is a random number drawn from a Gaussian distribution with a mean of zero and a standard deviation of unity.

{\bf 1.0$<$SNR$<$6.0: } \citet[][hereafter RP94]{1994A&A...287..676R} observed that measurements of weak lines (SNR$<$6) are strongly biased upwards, and the uncertainty on such lines cannot be well represented by a normal distribution.  We refer to this effect henceforth as the RP effect.  Taking the RP94 effect into account is straightforward with a Monte Carlo approach, requiring only that a log-normal distribution with appropriate mean and standard deviation is used for weak lines, instead of the normal distribution used for strong lines.

For lines with F/$\sigma<$6, we randomise the line fluxes using the following procedure: we first determine the log-normal distribution appropriate to the measured signal-to-noise ratio.  RP94 gave parameters for the log-normal distributions of F$_{obs}$/F$_{true}$, as a function of the observed SNR, and we fitted the following equations to the parameters in their Table 6:

\begin{equation}
\mu = \frac{0.0765957}{snr^2} + \frac{1.86037}{snr} - 0.309695
\end{equation}

\begin{equation}
\sigma = \frac{-1.11329}{snr^3} + \frac{1.8542}{snr^2} - \frac{0.288222}{snr} + 0.18018
\end{equation}

From the observed SNR we thus obtain the appropriate $\mu$ and $\sigma$.  Then, the line flux is found using the following equation:

\begin{equation}
F_i = \frac{F}{e^{(R\times\sigma + \mu)}}
\end{equation}

where R is a random number from a Gaussian distribution as before.  The result of this procedure is that weaker lines are drawn from log-normal distributions that peak below the observed value, with the effect increasing as SNR decreases.

{\bf SNR$<$1.0: } If the quoted uncertainty is larger than the actual line flux, the code assumes that the quote flux represents a 5$\sigma$ upper limit, and thus draws the randomized line flux from a folded Gaussian distribution, with $\mu$=0 and $\sigma$=0.2$\times$F.  The line flux is given by

\begin{equation}
F_i = abs(R)\times0.2F
\end{equation}

\begin{figure}
\includegraphics[width=0.47\textwidth]{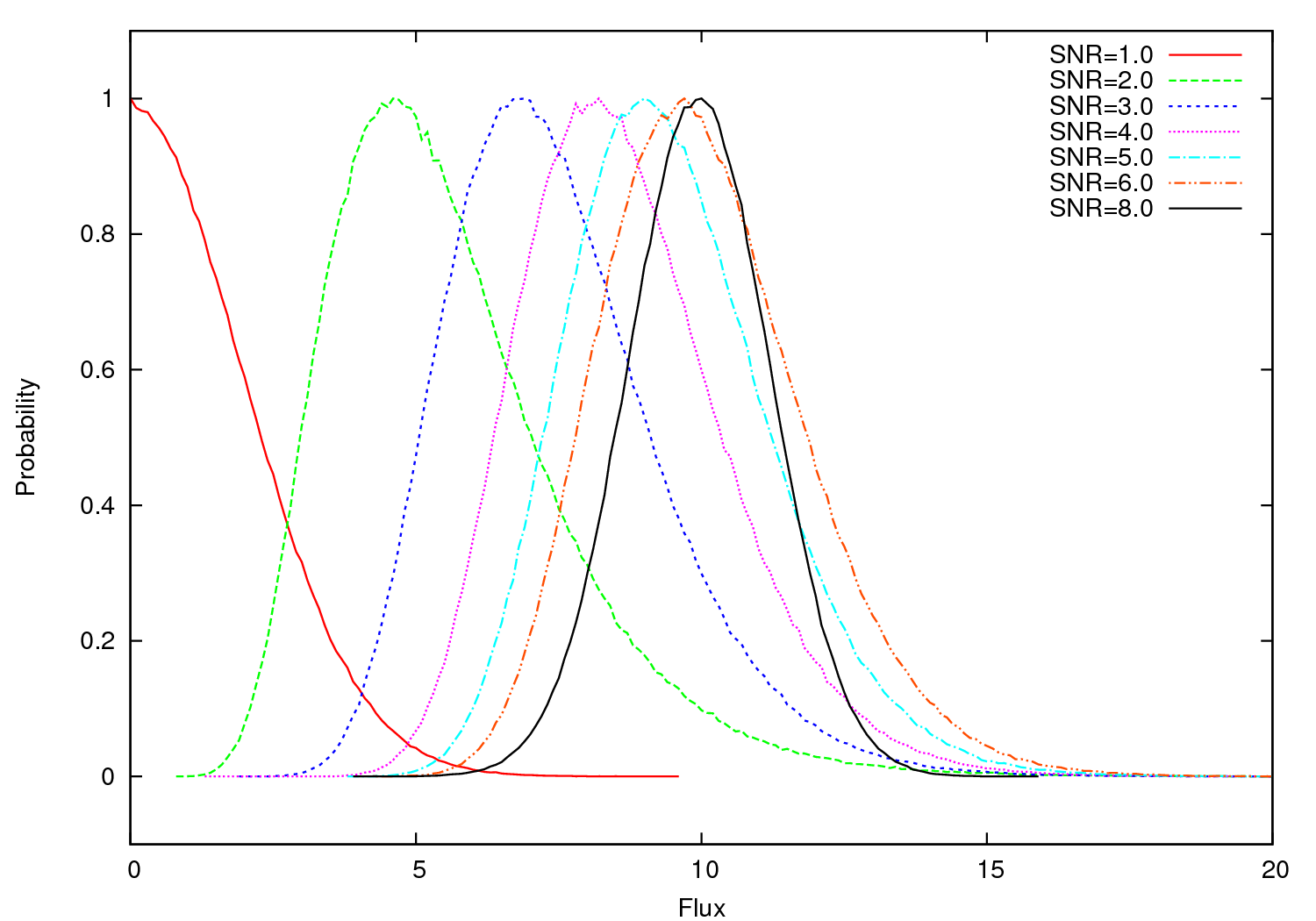}
\caption{The behaviour of {\sc neat}'s line flux randomization as a function of SNR, for a fixed measured flux of 10.0.  At high SNR (i.e. $>$6), normal distributions apply.  At lower SNR, the effect described by \citet{1994A&A...287..676R} results in log-normal distributions which are skewed towards lower values than the measured intensity.  If SNR$\leq$1 then the code assumes that the quoted flux is a 5$\sigma$ upper limit}
\label{distributions}
\end{figure}

Figure~\ref{distributions} shows the various possibilities that arise depending on the measured SNR.  The distributions plotted are from 10$^6$ runs of {\sc neat} in which 7 artificial line fluxes were randomized.  Each of the lines had a measured flux of 10.0, and the quoted uncertainties represented SNRs of 1.0, 2.0, 3.0, 4.0, 5.0, 6.0 and 8.0.  The figures shows how the normal distribution appropriate at high SNR is replaced by a log-normal distribution increasingly skewed towards lower values as SNR decreases.

\subsubsection{Random number algorithm}

\begin{figure}
\includegraphics[width=0.47\textwidth]{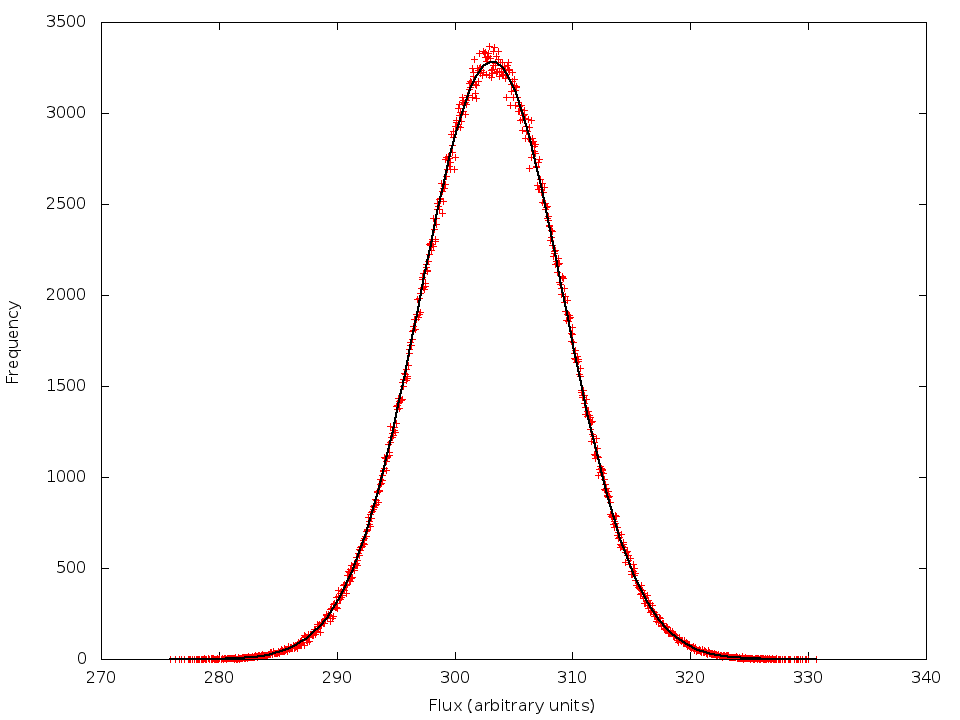}
\caption{A distribution of values produced by 1\,000\,000 runs of the random number generator, with the target Gaussian distribution overplotted.}
\label{gaussiantest}
\end{figure}

This process relies on the FORTRAN {\sc random\_number} command, seeded using the system clock and the time of the code's execution, to generate pseudo-random numbers.  The uniformly distributed numbers thus generated are then converted into a Gaussian distribution using an algorithm based on the ratio of uniforms method of \citet{Kinderman:1977:CGR:355744.355750}, available from \texttt{netlib.org}.  We tested the performance of this method by running the code 1\,000\,000 times, and plotting the distribution of fluxes obtained for an arbitrarily selected line.  We then fitted a Gaussian function to this distribution.  We found that the recovered mean was within 0.008\% of the specified value, while the recovered standard deviation was within 0.13\% of the specified value.  Figure~\ref{gaussiantest} shows the histogram of generated values with the required Gaussian distribution overplotted.  We thus consider that the random number generator in the code provides a reliably random Gaussian distribution.

\subsubsection{Sampling uncertainty}

The Monte Carlo approach relies on carrying out the analysis enough times to adequately sample the probability distributions of the derived quantities.  To determine what number of iterations suffices for this purpose, we ran the code 1,000,000 times, using emission line fluxes measured for IC1747 by \citet{2005MNRAS.362..424W}.  We then considered subsets of the output from this run.

To quantify the sampling uncertainty, we fitted a Gaussian to the observed probability distribution of the measured [O~{\sc iii}] temperatures, for each subset of iterations.  We chose this quantity as its actual uncertainty distribution should closely approximate a Gaussian.  The uncertainty on the Gaussian fit is thus a measure of how well the output distribution was sampled, for a given number of iterations.  A Gaussian output makes it straightforward to quantify the sampling uncertainty but the magnitude of this uncertainty is a function only of the number of iterations and not of the output distribution, and so the result applies generally.

\begin{figure*}
\includegraphics[width=0.48\textwidth]{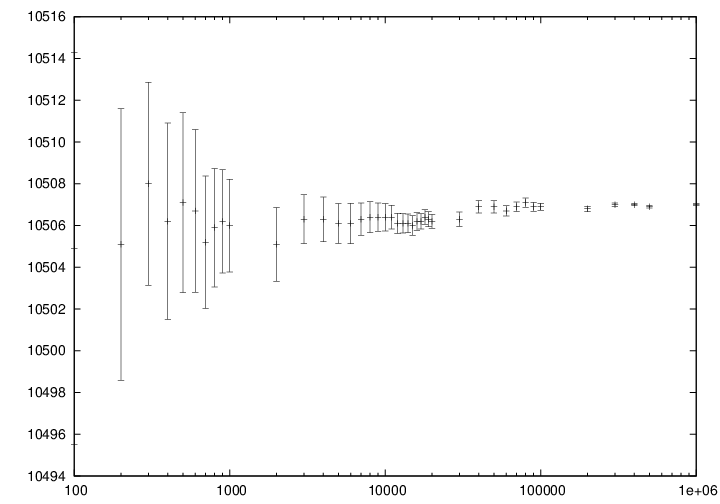}
\includegraphics[width=0.48\textwidth]{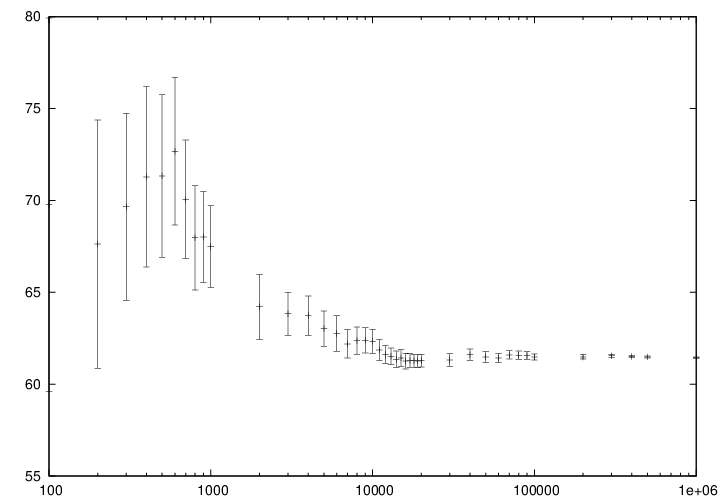}
\caption{The uncertainty on the parameters of gaussian fits to a {\sc neat} uncertainty distribution for T([O~{\sc iii}], as a function of the number of iterations.  The figures show the uncertainty on the fitted values of $\mu$ (l) and $\sigma$ (r).  We use these as a proxy for the Monte Carlo sampling uncertainty.}
\label{samplefigure1}
\end{figure*}

In Figure~\ref{samplefigure1} we show how the uncertainty of the fitted $\mu$ and $\sigma$ vary with the total number of iterations.  We find that the precision of the fit improves indefinitely with increasing number of iterations up to the limit of our investigation.  In our own investigations we carried out 10\,000 iterations on each line list we investigated; Figure~\ref{samplefigure1} shows that by this number of iterations, the sampling uncertainty on the [O~{\sc iii}] temperature is of the order of 1K.  We carried out our investigations on single processors of moderately powerful desktop and laptop machines, on which 10\,000 iterations typically took around 40-60 minutes.  We plan to parallelise {\sc neat} to enable larger numbers of iterations to be carried out in a conveniently short time.

\subsection{Interstellar extinction}
\label{extinctionsection}
The first step of any abundance analysis is a correction for interstellar extinction.  The amount of extinction is determined by {\sc neat} from the ratios of hydrogen Balmer lines, in an iterative procedure: the extinction is first calculated assuming intrinsic H$\alpha$, H$\beta$ and H$\gamma$ line ratios for a temperature of 10\,000\,K and a density of 1000\,cm$^{-3}$.  c(H$\beta$) is calculated from the flux-weighted average values derived from ratios of the three lines to their intrinsic values, and the line list is de-reddened using this value of c(H$\beta)$.  Temperatures and densities calculated as described below.  Then, the intrinsic Balmer line ratios are recalculated at the appropriate temperature and density, and c(H$\beta$) is recalculated.

The user can select the particular extinction law to be used.  Five extinction laws are currently available: the Galactic extinction curves of \citet{1983MNRAS.203..301H}, \citet{1990ApJS...72..163F} and \citet{1989ApJ...345..245C}, the Large Magellanic Cloud law of \citet{1983MNRAS.203..301H}, and the Small Magellanic Cloud law of \citet{1984A&A...132..389P}.  Adding further extinction laws would be straightforward, should the user wish to do so.

In section~\ref{extinction} we investigate the effect on derived quantities of the likely uncertainty in $R$, the ratio of selective to total extinction.

\subsection{Temperatures and densities}

Temperatures and densities are calculated using traditional collisionally excited line diagnostics.  For the purposes of subsequent abundance calculations, the nebula is divided into three ``zones", of low, medium and high excitation.  In each zone, temperatures and densities are calculated iteratively and weighted according to the reliability of each diagnostic.  Table~\ref{zonestable} shows the diagnostics used and the weighting given in each zone.

The scheme to calculate the densities is iterative, and proceeds as follows:

\begin{enumerate}
\item A temperature of 10,000K is initially assumed, and the density is then calculated from the line ratios relevant to the zone.
\item The temperature is then calculated from the temperature diagnostic line ratios, using the derived density.
\item The density is recalculated using the appropriately weighted average of the temperature diagnostics
\item The temperature is recalculated using this density.
\end{enumerate}

This iterative procedure is carried out successively for low, medium and high ionization zones, and in each case if no diagnostics are available, the temperature and/or density will be taken to be those derived for the previous zone.

\begin{table}
\begin{tabular}{ccc}
\hline
\multicolumn{3}{c}{Low ionization zone}\\
\hline
Diagnostic & Lines & Weight \\
~[O~{\sc ii}] density & $\lambda$3727/$\lambda$3729 & 1 \\
~[S~{\sc ii}] density & $\lambda$6717/$\lambda$6731 & 1 \\
~\\
~[N~{\sc ii}] temperature & $\frac{\lambda 6548 + \lambda 6584}{\lambda 5754}$ & 5 \\
~[S~{\sc ii}] temperature & $\frac{\lambda 6717 + \lambda 6731}{\lambda 4068 +  \lambda 4076}$ & 1 \\
~[O~{\sc ii}] temperature & $\frac{\lambda 7319,20 + \lambda 7330,31}{\lambda 3726 + \lambda 3729}$ & 1 \\
~[O~{\sc i}] temperature & $\frac{\lambda 6363 + \lambda 6300}{\lambda 5577}$ & 1 \\
~[C~{\sc i}] temperature & $\frac{\lambda 9850 + \lambda 9824}{\lambda 8727}$ & 1 \\
\hline
\multicolumn{3}{c}{Medium ionization zone}\\
\hline
Diagnostic & Lines & Weight \\
~[Cl~{\sc iii}] density & $\lambda$5517/$\lambda$5537 & 1 \\
~[Ar~{\sc iv}] density & $\lambda$4711/$\lambda$4740 & 1 \\
~[C~{\sc iii}] density & $\lambda$1907/$\lambda$1909 & 1 \\
~[O~{\sc iii}] density & $\lambda$52$\mu$m/$\lambda$88$\mu$m & 0 \\
~[Ar~{\sc iii}] density & $\lambda$9$\mu$m/$\lambda$21.8$\mu$m & 0 \\
~[S~{\sc iii}] density & $\lambda$18.7$\mu$m/$\lambda$33.8$\mu$m & 0 \\
~[Ne~{\sc iii}] density & $\lambda$15.5$\mu$m/$\lambda$36.0$\mu$m & 0 \\
~\\
~[O~{\sc iii}] temperature & $\frac{\lambda 4959 + \lambda 5007}{\lambda 4363}$ & 4\\
~[Ar~{\sc iii}] temperature & $\frac{\lambda 7135 + 7751}{5192}$ & 2\\
~[Ne~{\sc iii}] temperature & $\frac{\lambda 3868 + 3967}{3342}$ & 2\\
~[S~{\sc iii}] temperature & $\frac{\lambda 9069 + 9531}{6312}$ & 1\\
~[Ne~{\sc iii}] temperature & $\frac{\lambda 3868 + 3967}{15.5 \mu m}$ & 0\\
~[O~{\sc iii}] temperature & $\frac{\lambda 4959 + 5007}{52 \mu m}$ & 0\\
\hline
\multicolumn{3}{c}{High ionization zone}\\
\hline
Diagnostic & Lines & Weight \\
~[Ne~{\sc iv}] density & $\frac{\lambda2423}{\lambda2425}$ & 1 \\
~\\
~[Ar~{\sc v}] temperature & $\frac{\lambda 6435 + \lambda7005}{\lambda4625}$ & 1 \\
~[Ne~{\sc v}] temperature & $\frac{\lambda 3426 + \lambda3345}{\lambda2975}$ & 1 \\
\end{tabular}
\caption{Diagnostics used in the calculation of physical conditions.}
\label{zonestable}
\end{table}

\subsection{Ionic abundances}

Ionic abundances are calculated from collisionally excited lines using the temperature and density appropriate to their ionization potential.  Where several lines from a given ion are present, the ionic abundance adopted is found by averaging the abundances from each ion, weighting according to the observed intensity of the line.

Recombination lines are also used to derive ionic abundances for helium and heavier elements.  In deep spectra, many more recombination lines may be available than collisionally excited lines.  The code calculates the ionic abundance from each individual recombination line intensity using the atomic data listed in Table~\ref{RL_atomic_data}.  Then, to determine the ionic abundance to adopt, it first derives an ionic abundance for each individual multiplet from the multiplet's co-added intensity, and then averages the abundances derived for each multiplet to obtain the ionic abundance used in subsequent calculations.

\subsection{Total elemental abundances}

Generally, not all ionization stages of an ion that are actually present in a nebula will be detected, due to limited wavelength coverage and sensitivity.  Total elemental abundances must be estimated using ionization correction schemes, which are derived from photoionization models, or similarities in ionization potentials, or a combination of the two.

The code currently includes the ICF scheme of \citet{1994MNRAS.271..257K}.  We plan to incorporate further ICFs, and in a forthcoming paper we will compare the magnitude of the systematic uncertainties arising from the choice of ICF with the statistical uncertainties.

\subsection{Output}

By randomising and analysing the line list many times, it is possible to build up an accurate picture of the true distribution of statistical uncertainties associated with the chemical abundances and empirical diagnostics resulting from the line flux uncertainties. {\sc neat} collates all of the results from each iteration, and calculates uncertainties as follows: first of all for each output parameter it extracts the values containing 34.1\% of all results above and below the median.  The data is then binned using a bin size of 0.05 times the difference between these two values.  From the binned data, the mode of the probability distribution is obtained, and the code reports the final quantity and its uncertainties as this mode, and the values such that 68.2\% of all results lie within the range of uncertainties.

This approach does not assume any particular distribution of probabilities, but if the distribution is normal, log-normal or exponential-normal, then the uncertainties thus returned correspond to one standard deviation as normally defined.  We find that uncertainty distributions are not always characterised and may be bimodal or multimodal (see Section~\ref{natureofuncertainties}), and we therefore recommend that users directly inspect the probability distributions.  We provide with {\sc NEAT} a small shell script that produces plots for easy inspection.

\section{The nature of actual uncertainties}
\label{natureofuncertainties}

In this section we discuss the nature of the uncertainties revealed by the Monte Carlo technique.  We find that three distinct behaviours emerge for the uncertainties on abundances, depending on the overall depth of the line list being analysed.  In some cases, where all lines being analysed are well detected, the final uncertainties are close to symmetric and can be well approximated by Gaussian distributions.  However, in the nebulae that we have analysed, the final distributions are better described by log-normal distributions than by Gaussian.

In a minority of cases, very unusual uncertainty distributions emerge which cannot be sensibly fitted by analytic functions.  The origin of such distributions arises in some cases from the methodology adopted.  For example, if temperature diagnostic lines are weakly detected, then in some fraction of {\sc neat} iterations the temperature may be undefined, and thus taken as the default of 10\,000\,K.  In other iterations it may have a determined value different from 10\,000\,K.  The temperature distribution then becomes double peaked, and this propagates into CEL abundances, although it has little effect on RL abundances.  In situations like these, one can say that the true uncertainty distribution is broader even than {\sc neat} suggests, and must realistically encompass both values of the double peaked distribution.

At intermediate stages, different behaviour may be observed.  In particular, we find that for temperatures derived from collisionally excited lines, the uncertainty distribution is sometimes well characterised by an exponential-normal distribution; that is, the probability of e$^T$ is normally distributed.  In every case that we examined, though, the convolution of this distribution with the processes involved in calculating abundances resulted in a final abundance uncertainty distribution that was either normal or log-normal.

\begin{figure*}
\includegraphics[width=0.48\textwidth]{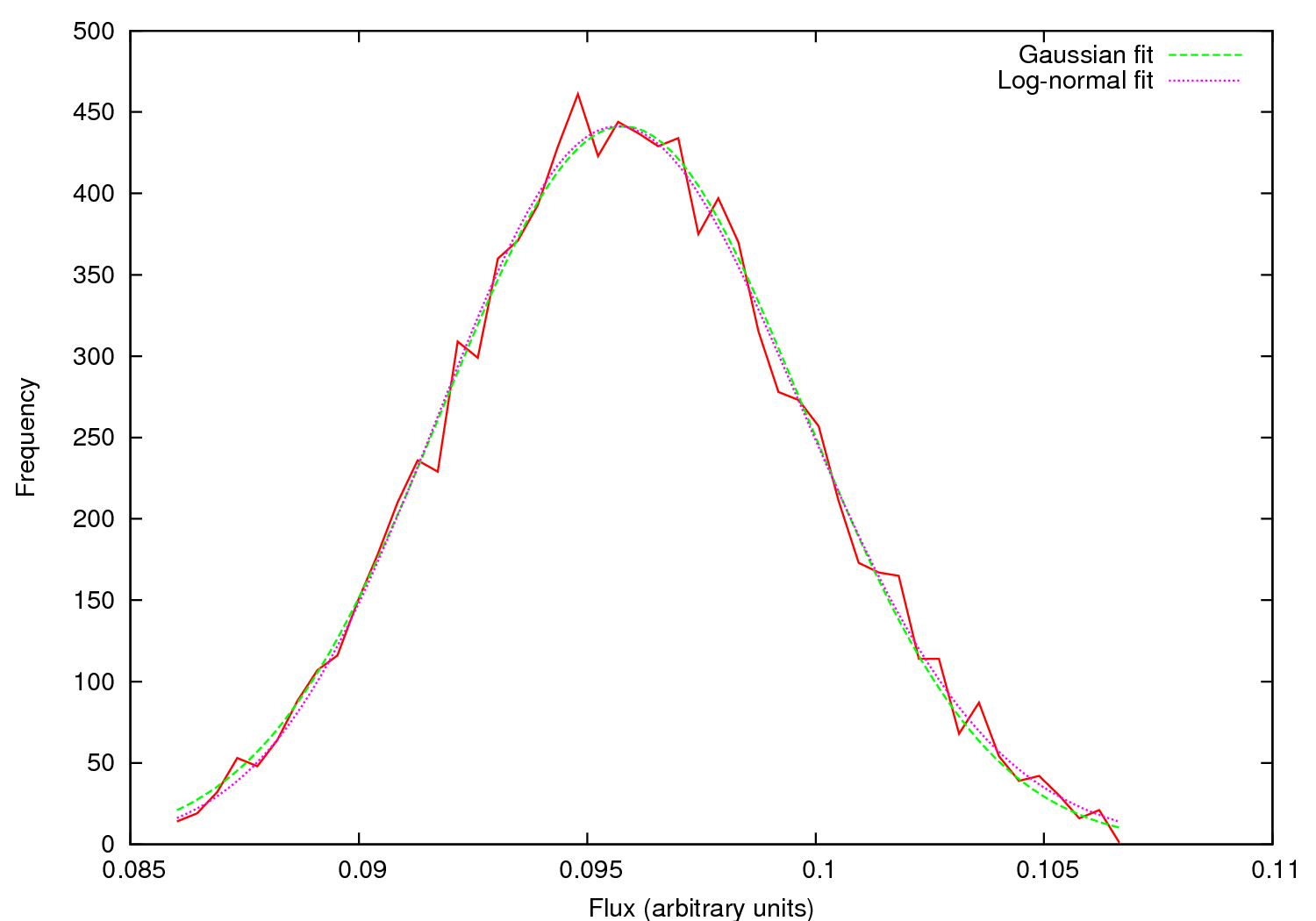}
\includegraphics[width=0.48\textwidth]{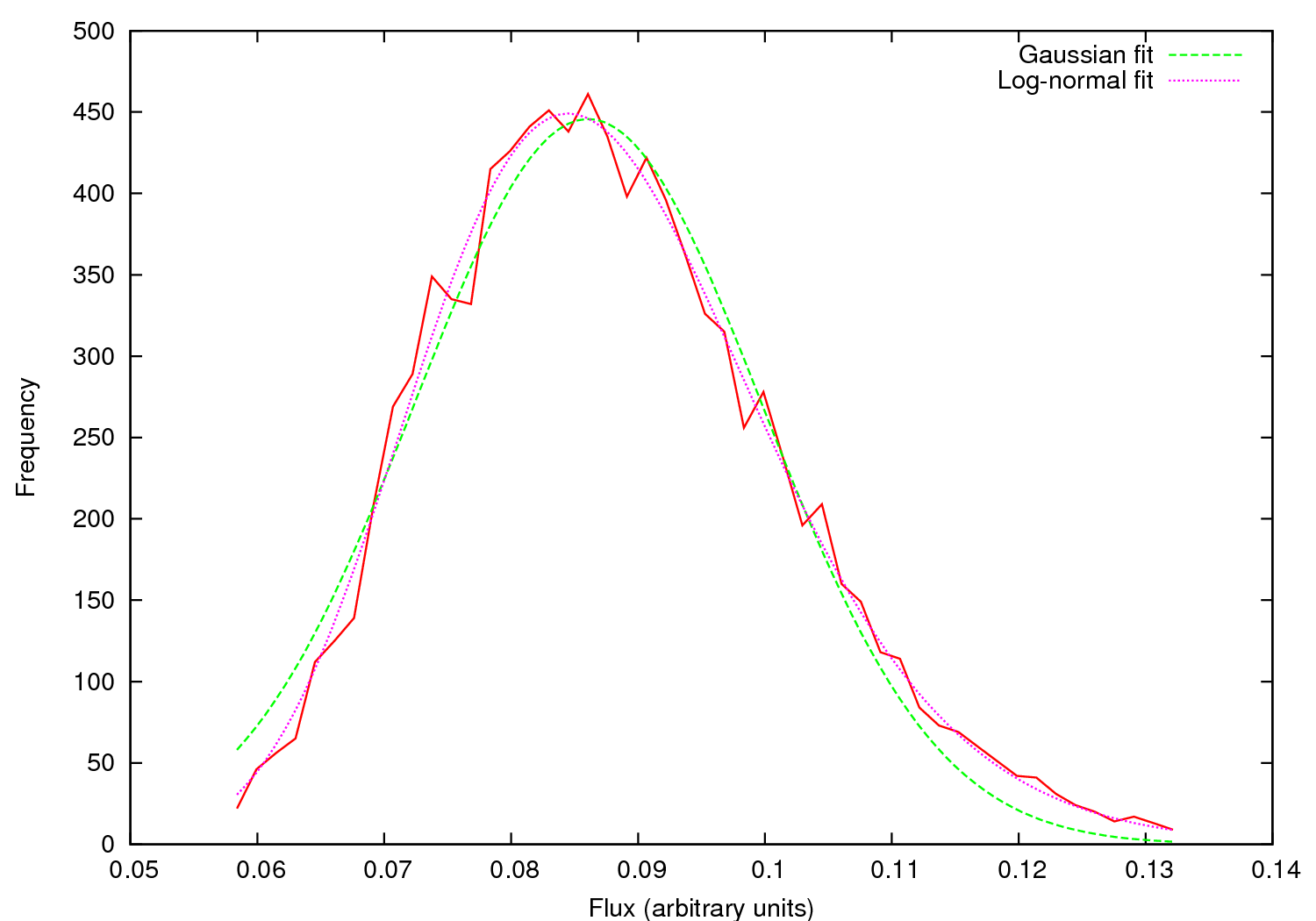}
\includegraphics[width=0.48\textwidth]{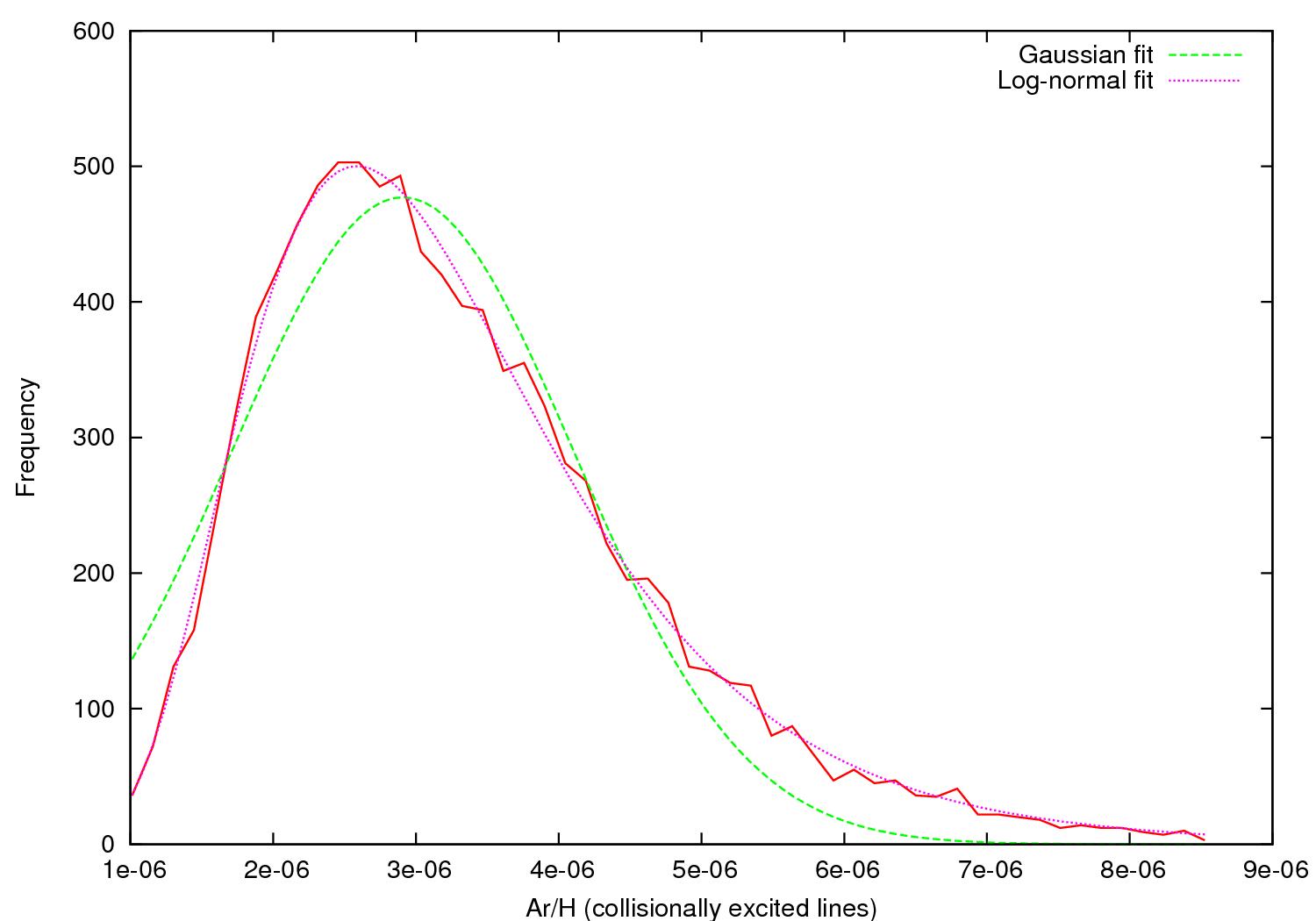}
\includegraphics[width=0.48\textwidth]{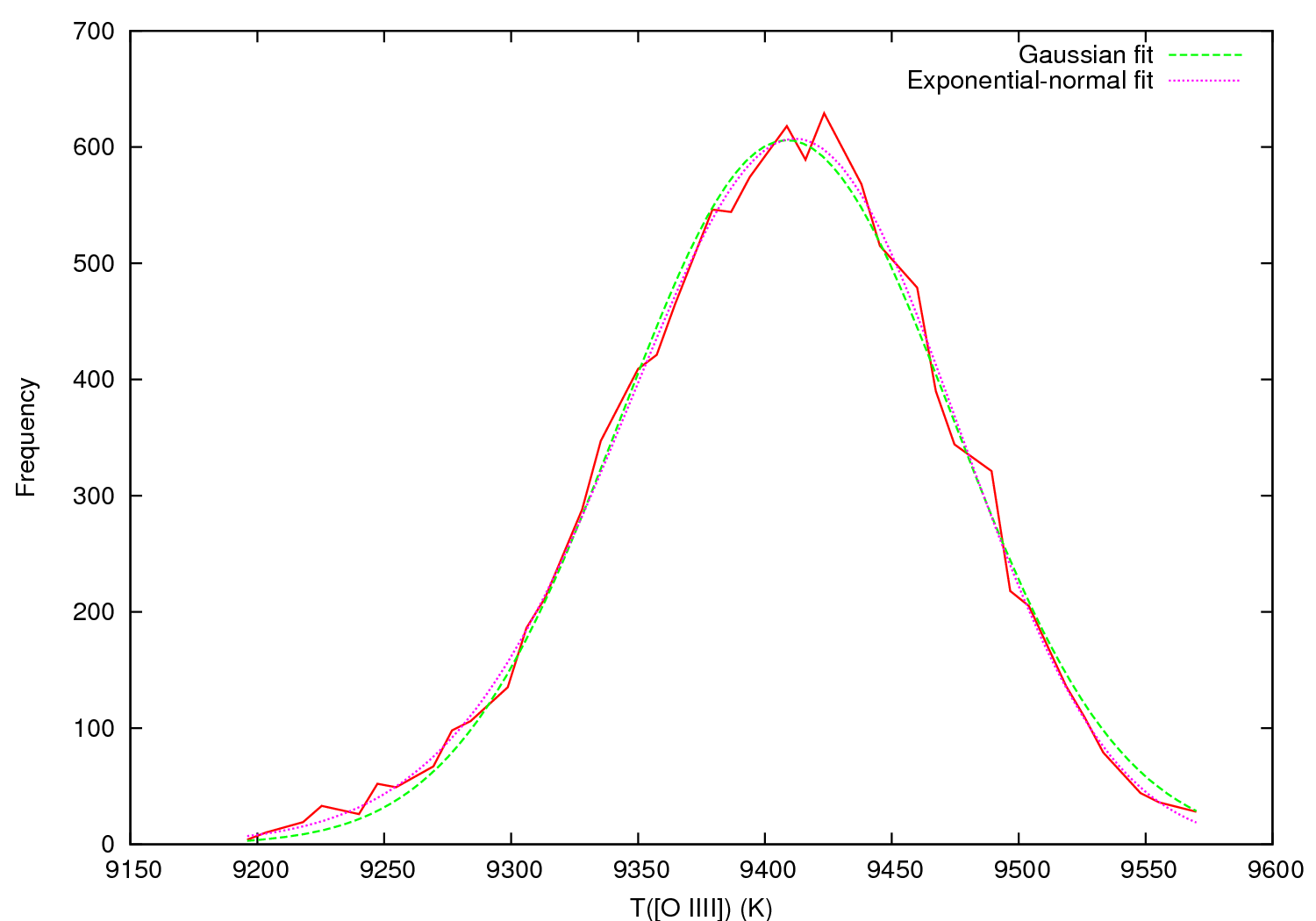}
\includegraphics[width=0.48\textwidth]{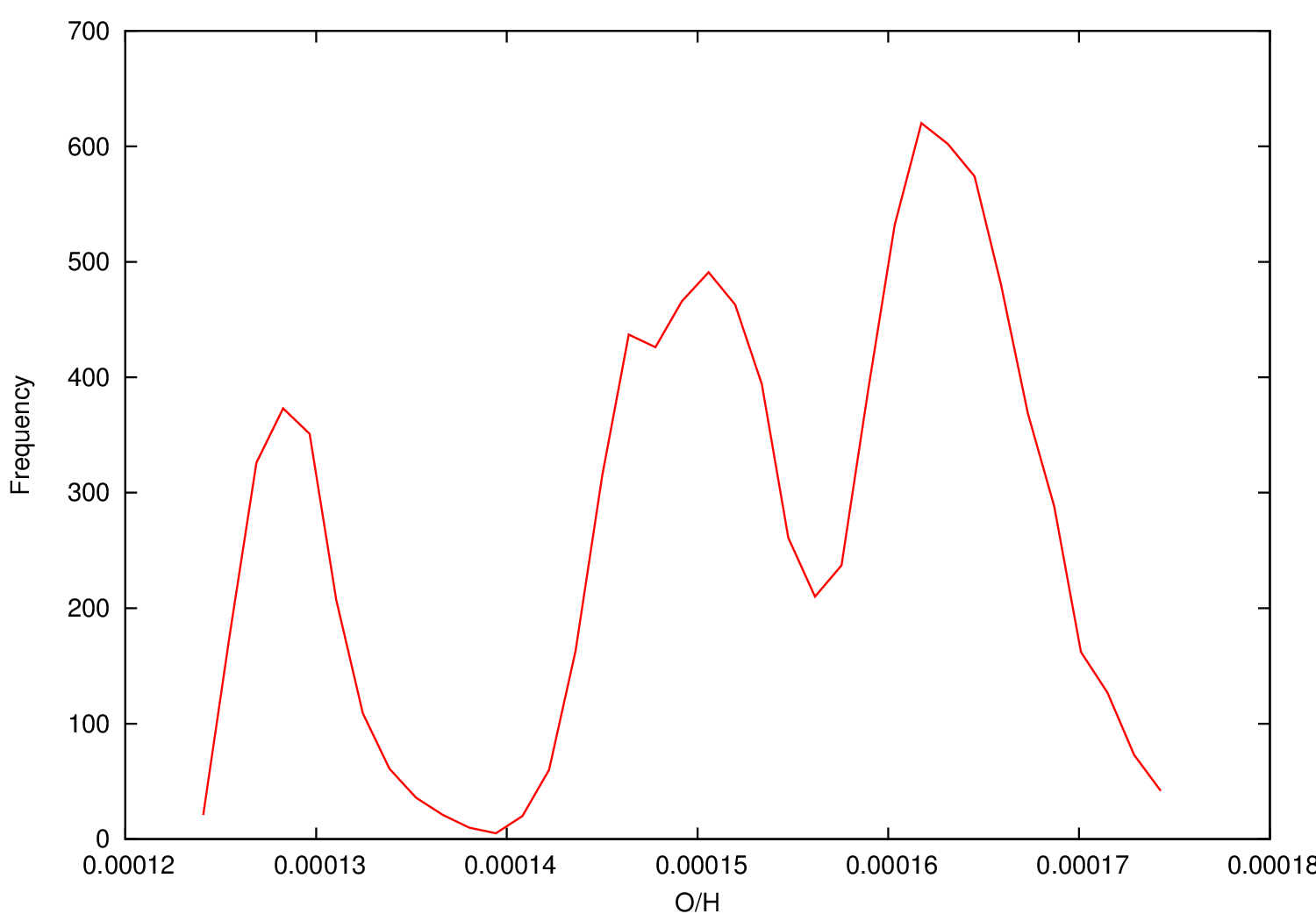}
\includegraphics[width=0.48\textwidth]{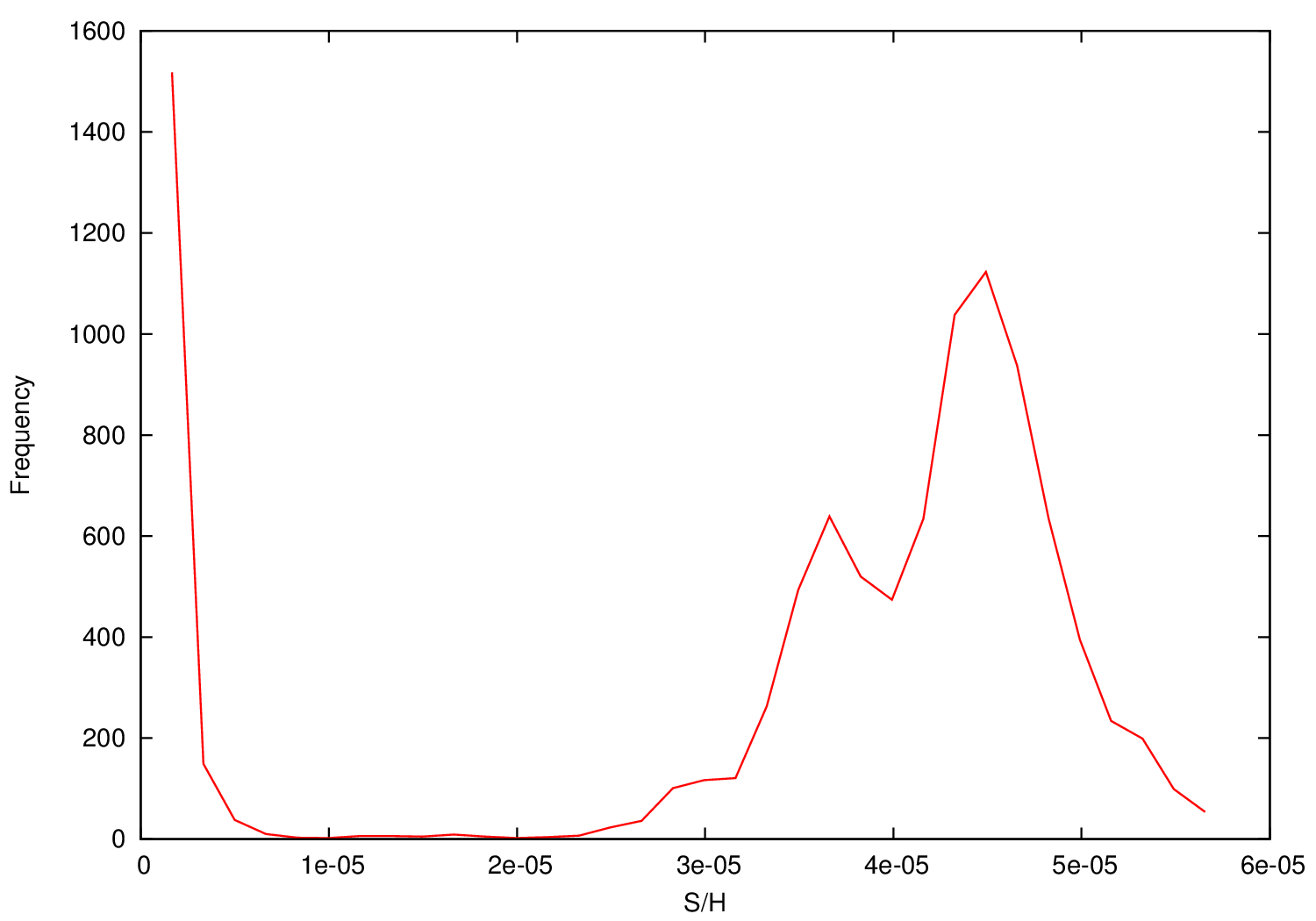}
\caption{Some representative examples of probability distributions emerging from this analysis.  These results are from analyses of the Orion Nebula (top left), WR nebula BAT99-11 (top right), Cn\,3-1 (middle left), NGC\,6803 (middle right), Sp 4-1 (bottom left) and DdDm 1 (bottom right).}
\label{Typical_uncertainties_images}
\end{figure*}

We show a selection of illustrative examples in Figure~\ref{Typical_uncertainties_images}, and we plot Gaussian, log-normal and exponential-normal fits to the distributions shown where possible.  In Table~\ref{Typical_uncertainties_table} we give the parameters of the fits to the plotted distributions.  The RMS of the residuals is given as a quantitative measure of which distribution is a better fit to the data.

\begin{table*}
\begin{tabular}{lllllllll}
\hline
Nebula & Quantity & \multicolumn{3}{c}{Gaussian} & \multicolumn{3}{c}{Log-normal} & Reference \\
       &          & $\mu$ & $\sigma$ & RMS of residuals & $\mu$ & $\sigma$ & RMS of residuals\\
\hline
%NGC 6543 & He/H     & 0.117 & 0.003 & 15.5 & -0.930 & 0.023 & 14.7 & (1) \\
Orion    & He/H     & 0.096             & 0.004 & 14.0 & -1.019 & 0.041 & 13.5 & (1) \\
BAT99-11 & He/H     & 0.086             & 0.014 & 23.8 & -1.073 & 0.159 & 14.7 & (2) \\
Cn\,3-1  & Ar/H     & 2.91$\times$10$^{-6}$ & 1.20$\times$10$^{-6}$ & 41.1 & -12.87 & 0.410 & 11.1 & (3) \\
       &          & \multicolumn{3}{c}{Gaussian} & \multicolumn{3}{c}{Exponential-normal} &           \\
       &          & $\mu$ & $\sigma$ & RMS of residuals & $\mu$ & $\sigma$ & RMS of residuals\\
NGC\,6803 & T([O~{\sc iii}]) & 9410 & 65 & 18.4 & 12231.8 & 795.0 & 15.5 & (3)\\
%Sp 4-1   & O/H (RL) & 
%NGC 2022 & He/H     & 0.096 & 0.014 & 18.6 & -1.025 & 0.150 & 11.9 & (4) \\
\hline
\end{tabular}
\caption{Analytic fits to the uncertainty distributions shown in Figure~\ref{Typical_uncertainties_images}.  Log-normal distributions emerge much more frequently than Gaussian or unquantifiable distributions.  The line lists analysed are from (1) Esteban et al. (2004), (2) Stock, Barlow \& Wesson (2011), (3) Wesson, Liu \& Barlow (2005).  The exponential-normal parameters for NGC\,6803 are those resulting from fitting the uncertainty distribution with a function in which exp(T$_e$/1000) was normally distributed.}
\label{Typical_uncertainties_table}
\end{table*}

The finding that the majority of uncertainty distributions are best described by log-normal distributions demonstrates that analytic uncertainty propagation generally does not accurately quantify the true uncertainties arising from real measurements of the spectra of astronomical objects.  It also implies that temperatures, densities and abundances are generally more likely to be underestimated than overestimated, although this effect will generally be small.  The emergence of unquantifiable distributions shows that analytic techniques can break down very severely; most significantly there is on the face of it no obvious difference between the line lists that give generally log-normal final uncertainties, and those that give unquantifiable distributions.  Thus, it appears that there is no {\it a priori} way of telling whether the uncertainties are going to be well behaved or erratic.

\section{The RP effect}
\label{RPeffect}

As discussed earlier, {\sc neat} accounts for the RP effect, in which fluxes measured from lines with SNR$<$6 are generally overestimated, with the magnitude of the effect increasing as SNR$\to$1.  In this section we show the importance of this effect and demonstrate that flawed results will inevitably result if the effect is ignored.

To investigate the magnitude of this effect, we reanalysed two line lists - spectra of NGC\,6543, the Cat's Eye Nebula, presented by \citet{2004MNRAS.351.1026W}, and spectra of the Orion Nebula presented by \citet{2004MNRAS.355..229E}.  In both cases we ran two instances of {\sc neat}, one in which the RP effect was ignored, and all line flux uncertainties quoted in the two papers were assumed to represent Gaussian probability distributions, and the second in which the RP effect was accounted for as described in Section~\ref{randomising}.

One important and as yet unresolved issue in nebular abundance studies is the so-called abundance discrepancy problem (see for example \citet{2006IAUS..234..219L} for a review).  The RP effect can cause errors in the assessment of the magnitude of the discrepancy; recombination lines of heavy elements are much weaker than the collisionally excited lines of the same species, and are thus measured with lower signal to noise ratios.  In almost any real astronomical spectra, regardless of the number of lines detected, the weakest lines measured will be subject to the RP effect, and for deep spectra of photoionized nebulae, the weakest lines will almost all be recombination lines.  Thus, recombination line abundance measurements may be subject to an upward bias that collisionally excited line abundances are largely free from.

\begin{figure*}
\includegraphics[width=0.48\textwidth]{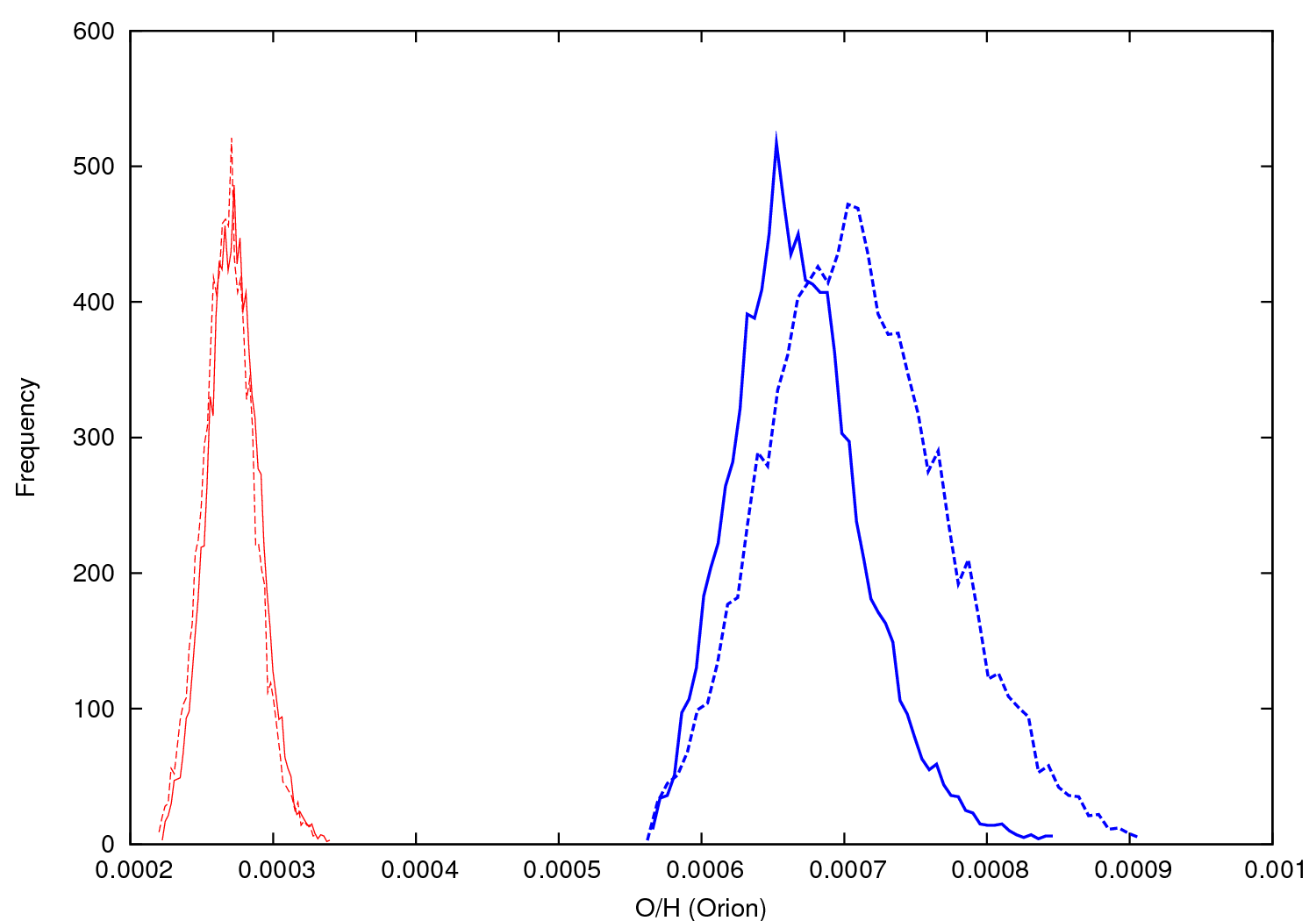}
\includegraphics[width=0.48\textwidth]{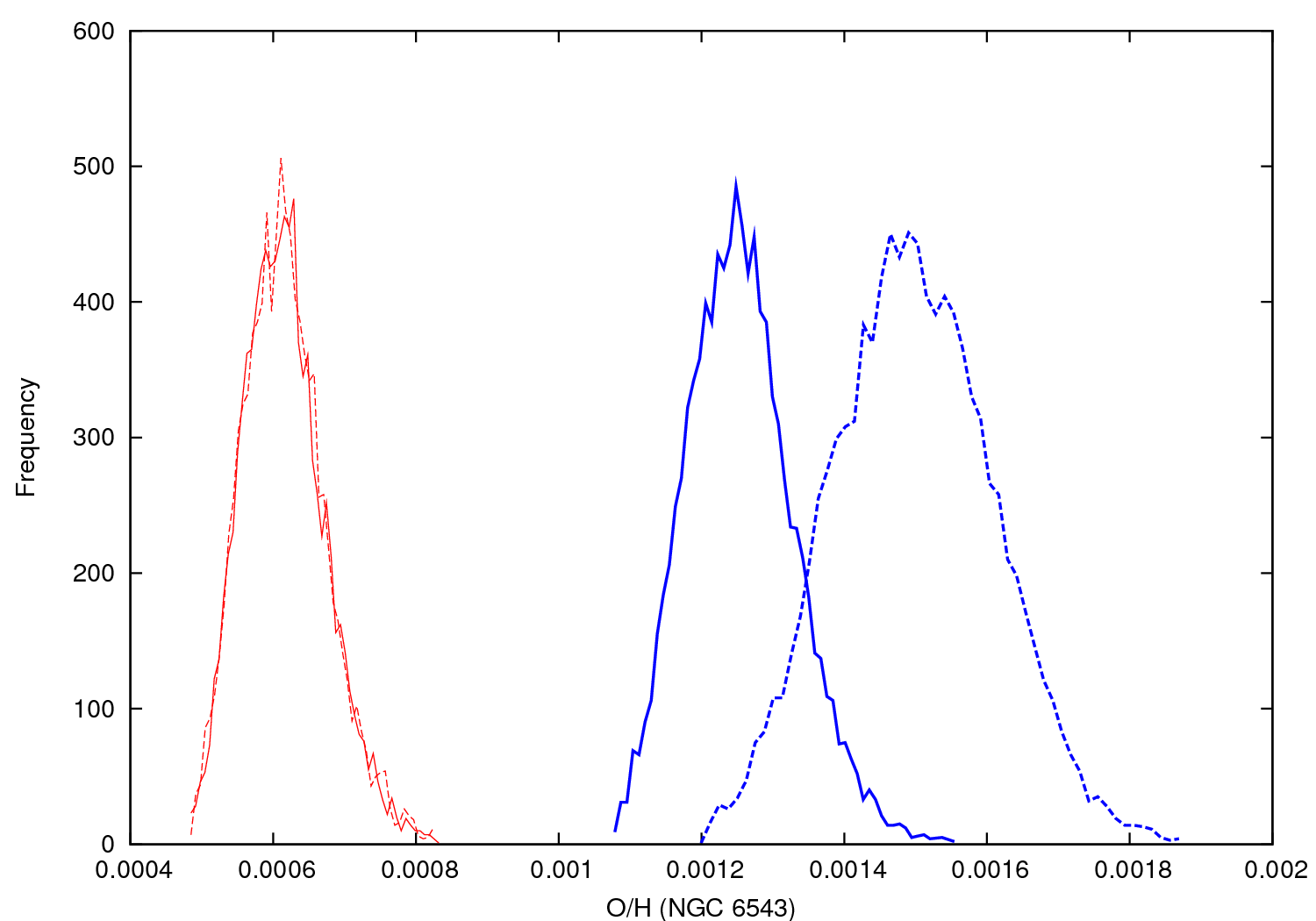}
\includegraphics[width=0.48\textwidth]{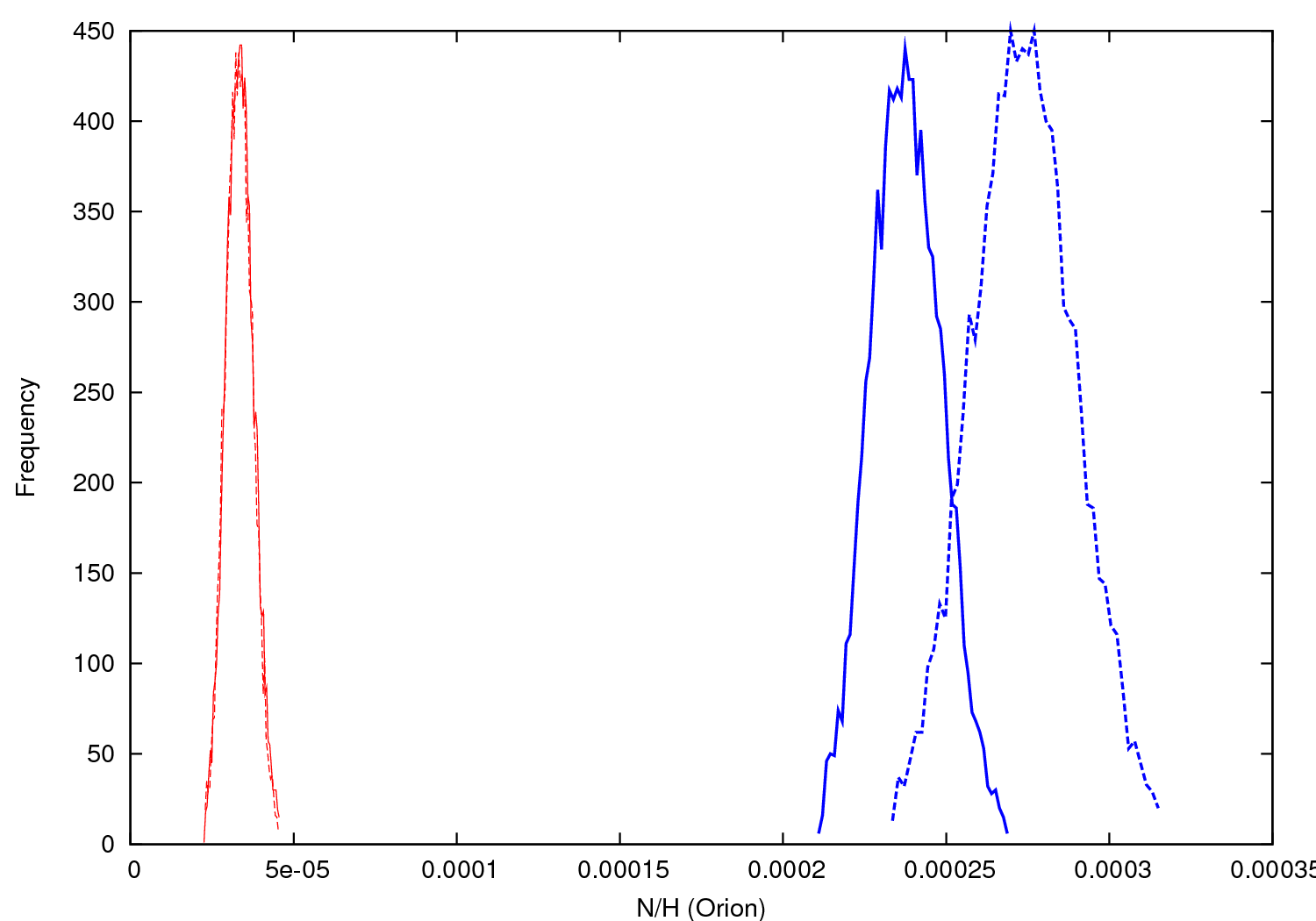}
\includegraphics[width=0.48\textwidth]{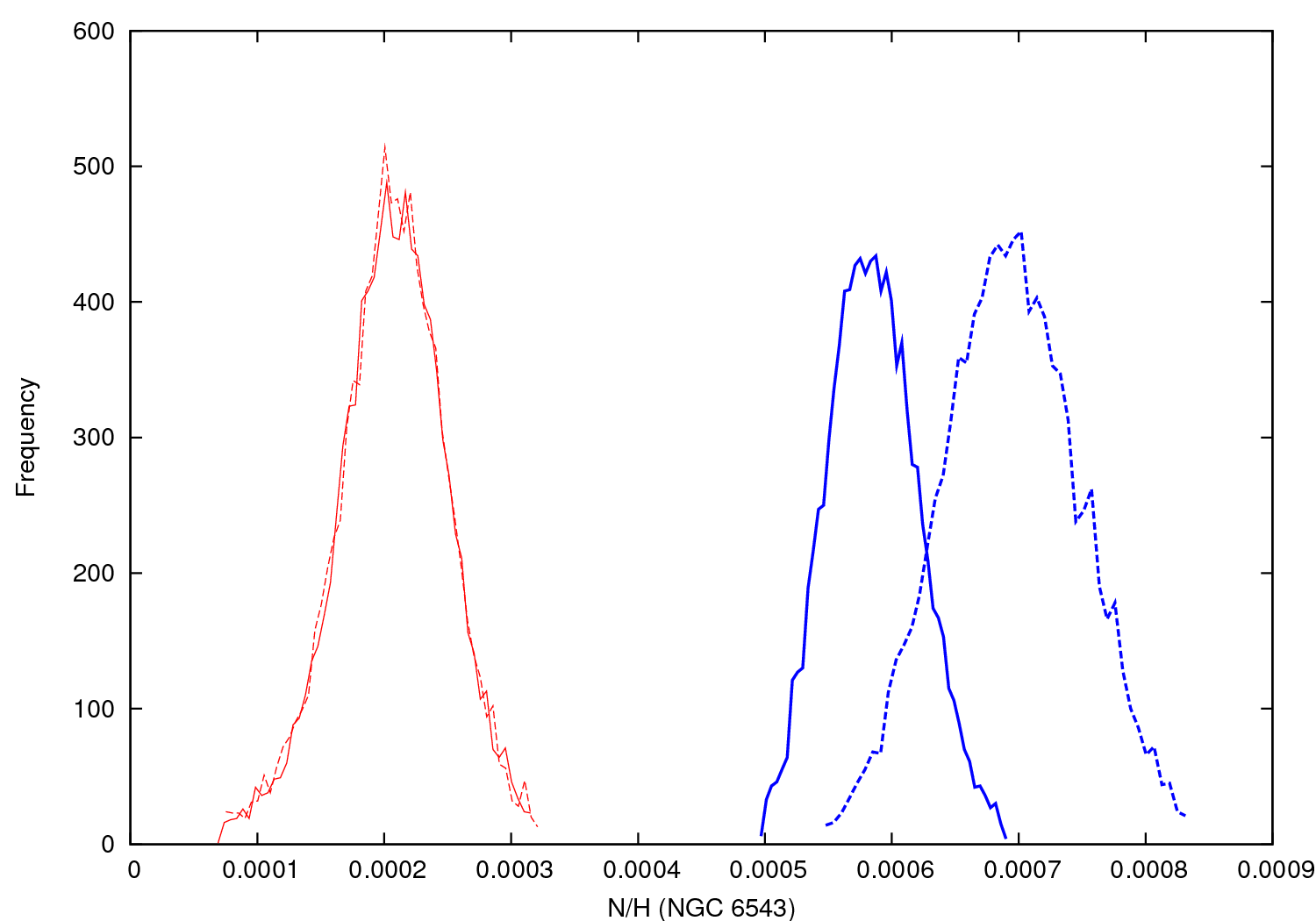}
\includegraphics[width=0.48\textwidth]{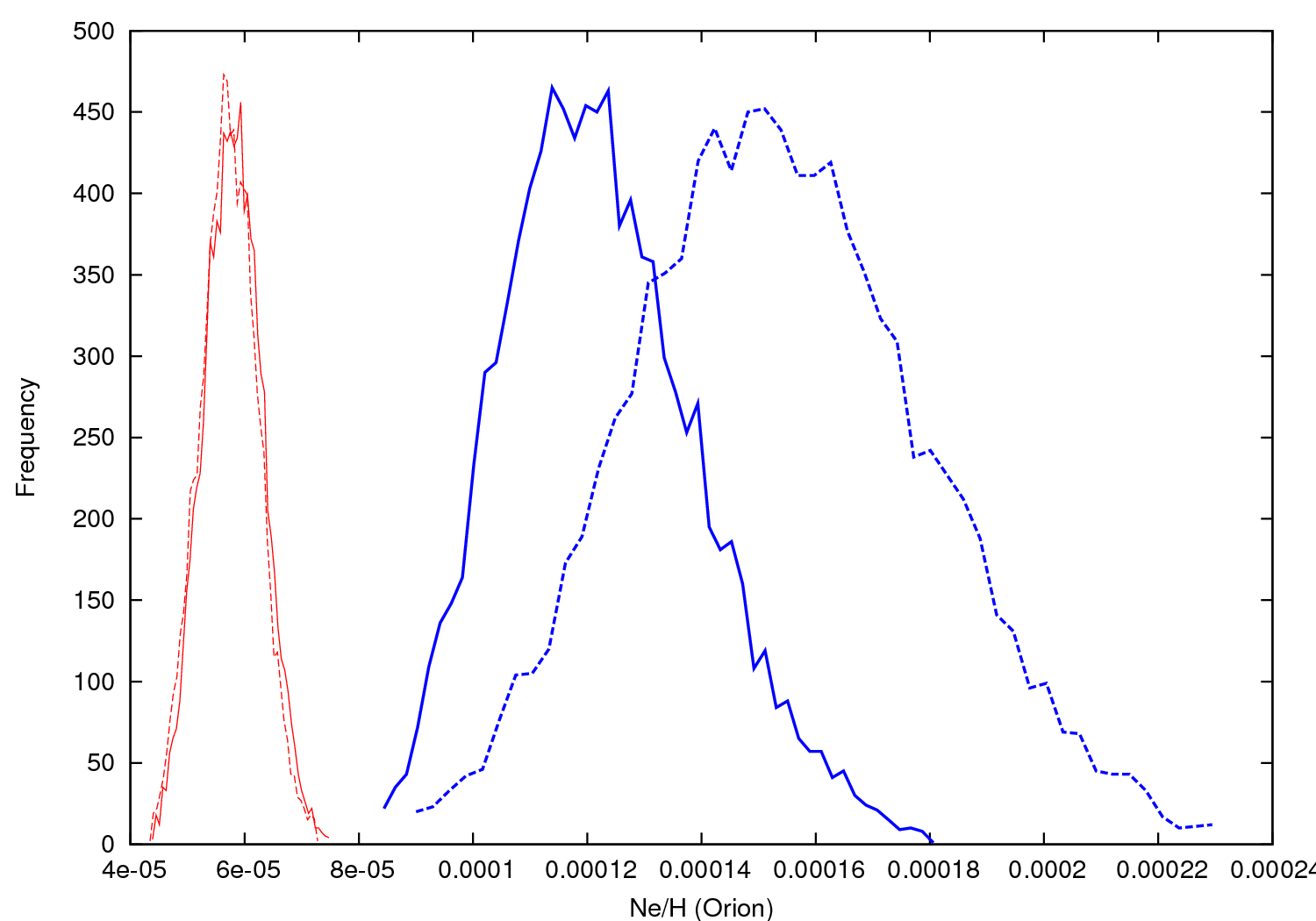}
\includegraphics[width=0.48\textwidth]{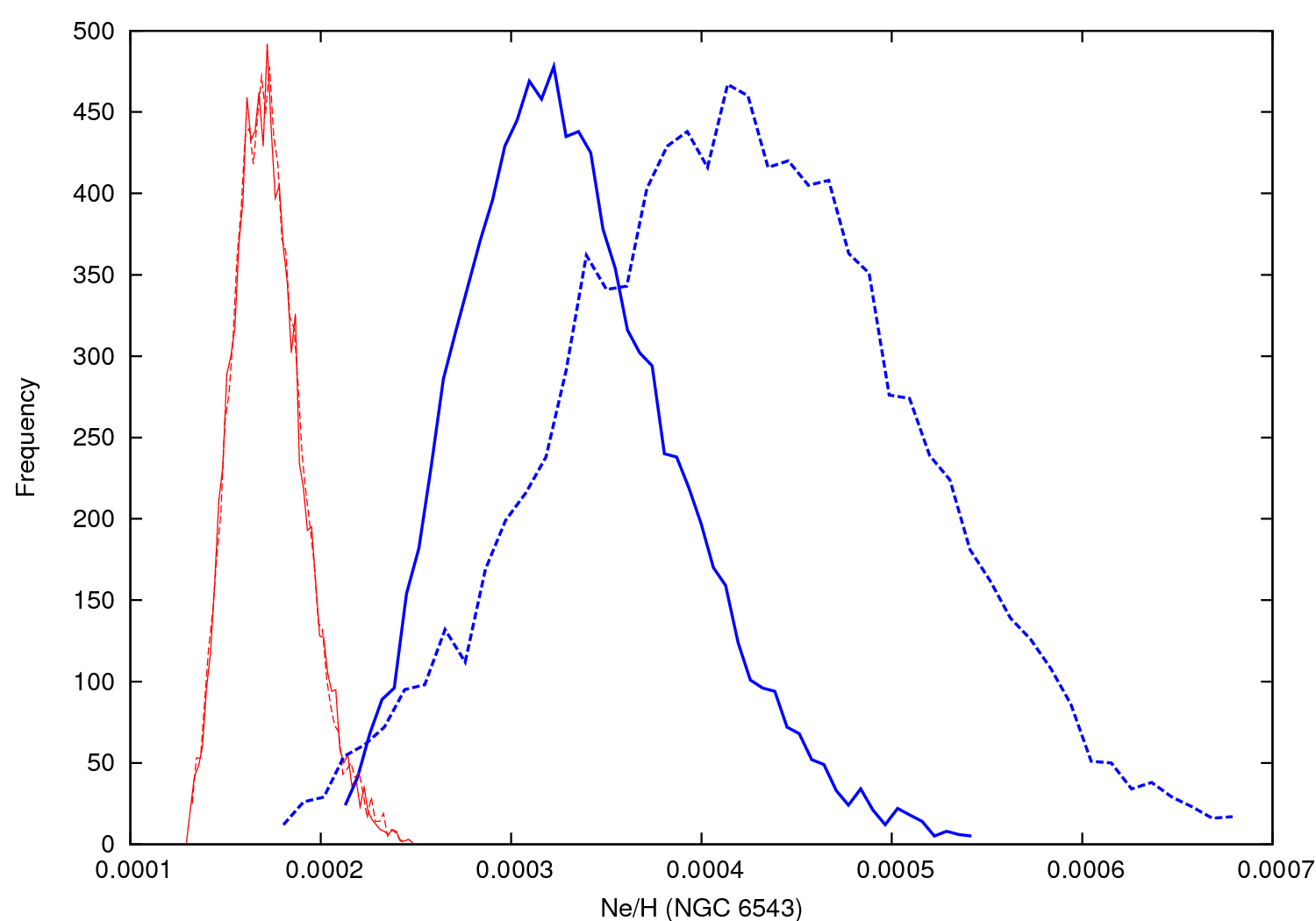}
\caption{The RP effect on abundance determinations.  Dashed lines show abundances derived assuming that weak line fluxes have a gaussian uncertainty distribution.  Solid lines show the results obtained by more realistically assuming a log-normal distribution as described in Section~\ref{RPeffect}.  Thin red and thick blue lines correspond to collisionally excited line and recombination line abundances respectively.  Only the far weaker RLs are subject to the RP effect.}
\label{RP_figures}
\end{figure*}

Figure~\ref{RP_figures} shows that in these two nebulae, ignoring the RP effect indeed has no effect on abundances of nitrogen, oxygen and neon derived from collisionally excited lines, but in all cases leads to an overestimate of the abundances derived from the recombination lines of these elements.  Furthermore, it turns out that properly accounting for the non-Gaussian nature of the uncertainties on weak lines leads to a significant reduction in the uncertainty associated with abundances determined from them; we fit the resulting uncertainty histograms with Gaussian functions, and find that accounting for the RP effect reduces the relative standard deviation on the measured abundances by 25-30\%.  Table~\ref{RP_numbers} summarises the mean and standard deviation of the final abundance determinations from recombination lines.

\begin{table*}
\begin{tabular}{l ll l}
\hline
Element & 10$^4 \times$ X/H (RP ignored) & 10$^4 \times$ X/H (RP included) & $\frac{\sigma/\mu (RP included)}{\sigma/\mu (RP ignored)}$ \\
\hline
\multicolumn{4}{c}{Orion}\\

N & 2.74 $\pm$ 0.16 & 2.37 $\pm$ 0.11 & 0.79 \\
O & 7.04 $\pm$ 0.63 & 6.63 $\pm$ 0.43 & 0.73 \\
Ne & 1.52 $\pm$ 0.26 & 1.20 $\pm$ 0.17 & 0.83 \\

\multicolumn{4}{c}{NGC 6543} \\

N & 6.93 $\pm$  0.56 & 5.83 $\pm$  0.37 & 0.78 \\
O & 14.94 $\pm$ 1.15 & 12.48 $\pm$ 0.74 & 0.77 \\
Ne & 4.17 $\pm$ 0.94 & 3.23 $\pm$ 0.56 & 0.77 \\
\hline
\end{tabular}
\caption{Results of heavy element abundance determinations from recombination lines, using the published line lists of \citet{2004MNRAS.351.1026W} and \citet{2004MNRAS.355..229E} for the Orion Nebula.}
\label{RP_numbers}
\end{table*}

In other shallower spectra, it may often be the case that the auroral lines [N~{\sc ii}] $\lambda$5754 and [O~{\sc iii}] $\lambda$4363 are measured with low enough SNR that they become subject to the RP effect.  In this case, the derived temperatures would be overestimated, and collisionally excited line abundances underestimated.  To investigate this effect, we analysed the line list of H1013, an extragalactic H~{\sc ii} region in the spiral galaxy M101, presented by \citet{2009ApJ...700..654E}.  In this nebula, the [O~{\sc iii}] line at 4363{\AA} is detected with an SNR of only 2.7.

\begin{figure*}
\includegraphics[width=0.48\textwidth]{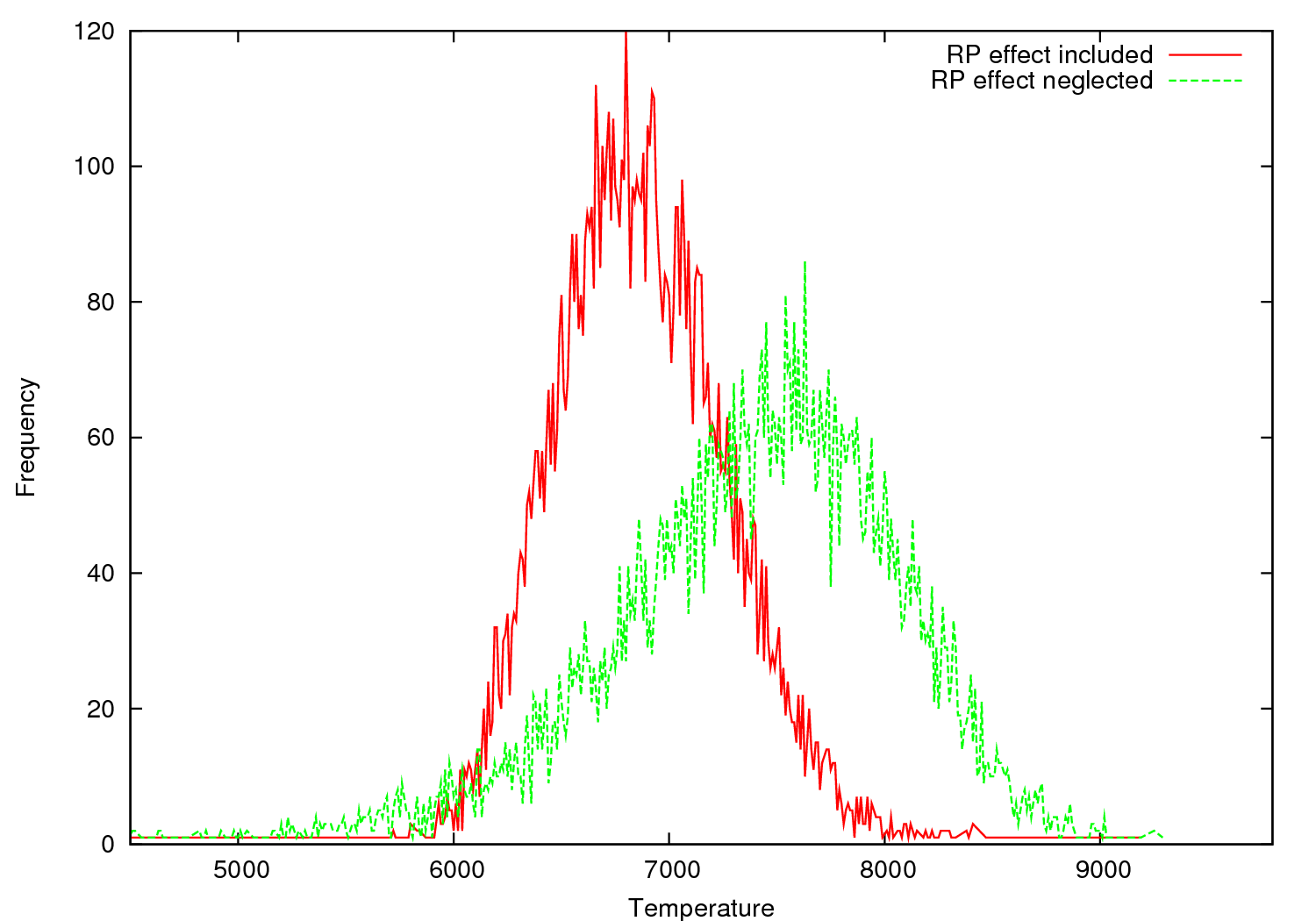}
\includegraphics[width=0.48\textwidth]{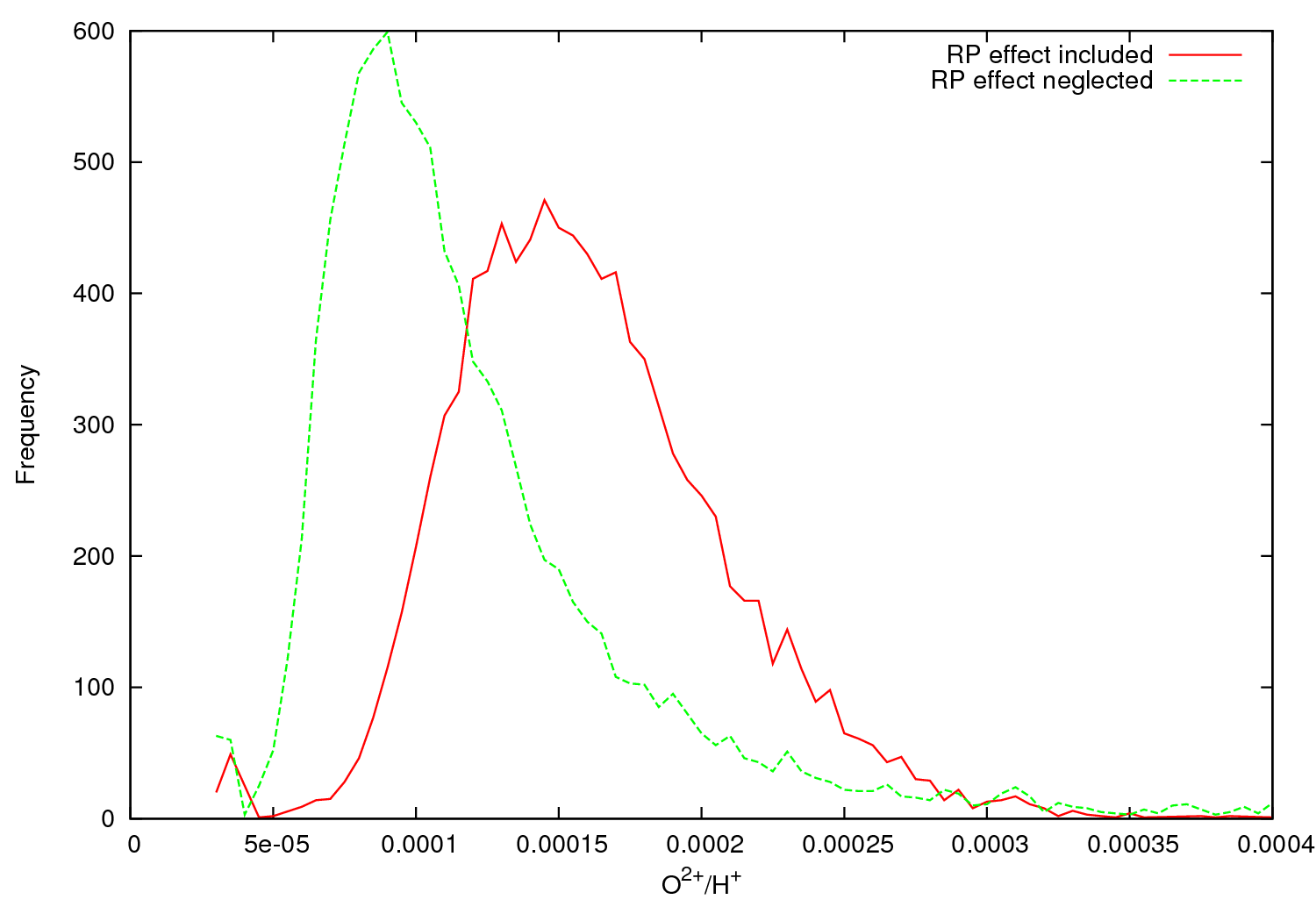}
\caption{The RP effect when CEL temperature diagnostic lines are measured with low SNR.  The figure shows a reanalysis of data for the H~{\sc ii} region H1013, showing the RP effect on the derived [O~{\sc iii}] temperature (l) and O$^{2+}$/H$^+$ abundance (r).}
\label{h1013_RP_effect}
\end{figure*}

Figure~\ref{h1013_RP_effect} shows that significant systematic uncertainties are produces when the RP effect on temperature diagnostic lines is ignored.  When the effect is neglected, we determine an [O~{\sc iii}] temperature of 7480$\pm$610\,K, in very close agreement with the value of 7370$\pm$630\,K reported by \citet{2009ApJ...700..654E}.  However, when we account for the RP effect, we find a value of 6840$\pm$390\,K.  Similarly for abundances, considering O$^{2+}$/H$^+$, we find that by neglecting the RP effect we obtain a value of 7.95$\pm$0.15 (on a logarithmic scale where N(H)=12), close to the value of 8.05$\pm$0.12 obtained by \citet{2009ApJ...700..654E}.  Accounting for the effect yields a value of 8.16$\pm$0.12.

The reduction in uncertainty when accounting for the RP effect arises in two distinct ways; firstly, as can be seen in Figure~\ref{h1013_RP_effect}, the probability distribution for the [O~{\sc iii}] temperature when the RP effect is not accounted for is neither normal nor log-normal but is instead better described by an exponential-normal distribution.  When the input uncertainty distribution is correctly characterised as a log-normal distribution, the convolution of this distribution with the processes which give rise to the exponential-normal distribution of temperature probabilities results in a final distribution which is narrower than when the input distribution is assumed to be normal.

The second effect is that when abundances from many weak lines are being combined to derive an abundance, accounting for the RP effect results in a modest reduction in the scatter of the derived abundances, and a corresponding reduction in the overall uncertainty on the combined abundance.

We re-emphasise, therefore, that neglecting this effect results in incorrect abundances.  Accounting for it removes a source of systematic uncertainty, and reduces the statistical uncertainty of the abundances determined.

One can see in Figure~\ref{RP_figures} that abundances derived from recombination lines have similar or smaller uncertainty distributions than those derived from collisionally excited lines, even though the line fluxes may be several orders of magnitude lower.  This reflects their very weak dependence on temperature and density; the uncertainties on the adopted temperatures and densities hardly propagate into RL abundances but have a significant effect on the CEL abundances.

\section{Interstellar extinction}
\label{extinction}

{\sc neat} allows a robust propagation of uncertainties from line flux measurements into derived quantities.  It is also possible to investigate the effect of statistical uncertainties arising at different stages of the process.  In this section, we consider the effect of the uncertainty in R, the ratio of total to selective extinction given by

\begin{equation}
R = \frac{A(V)}{E(B-V)}
\end{equation}

It is well known that R varies along different sight lines (eg \citet{2004ApJ...616..912V}, \citet{2005ApJ...623..897L}), but determining its value for particular objects is generally impractical and instead, it is commonly assumed to equal 3.1.  We investigate the effect of an uncertainty in this value by comparing analyses in which R is fixed to be 3.1, and in which R is drawn from a Gaussian distribution with $\mu$=3.1 and $\sigma$=0.15.  For this investigation we used the R-dependent extinction law parametrization of \citet{1989ApJ...345..245C}.  We took the uncertainty as the largest value that we found quoted in the literature for the diffuse ISM \citep{2005ApJ...623..897L}.

We analysed the emission line measurements of NGC 7026 presented by \citet{2005MNRAS.362..424W}, including the line flux uncertainties which were not published in that paper.  We chose this object as it is significantly reddened (c(H$\beta$)=1.0), and has many very well detected lines in its spectrum (142 lines measured, 120 with SNR$>$3 and 65 with SNR$>$10).  This combination should maximise the effect of an uncertainty on R, relative to the effect of the line flux measurement uncertainties.

\begin{figure*}
\includegraphics[width=0.48\textwidth]{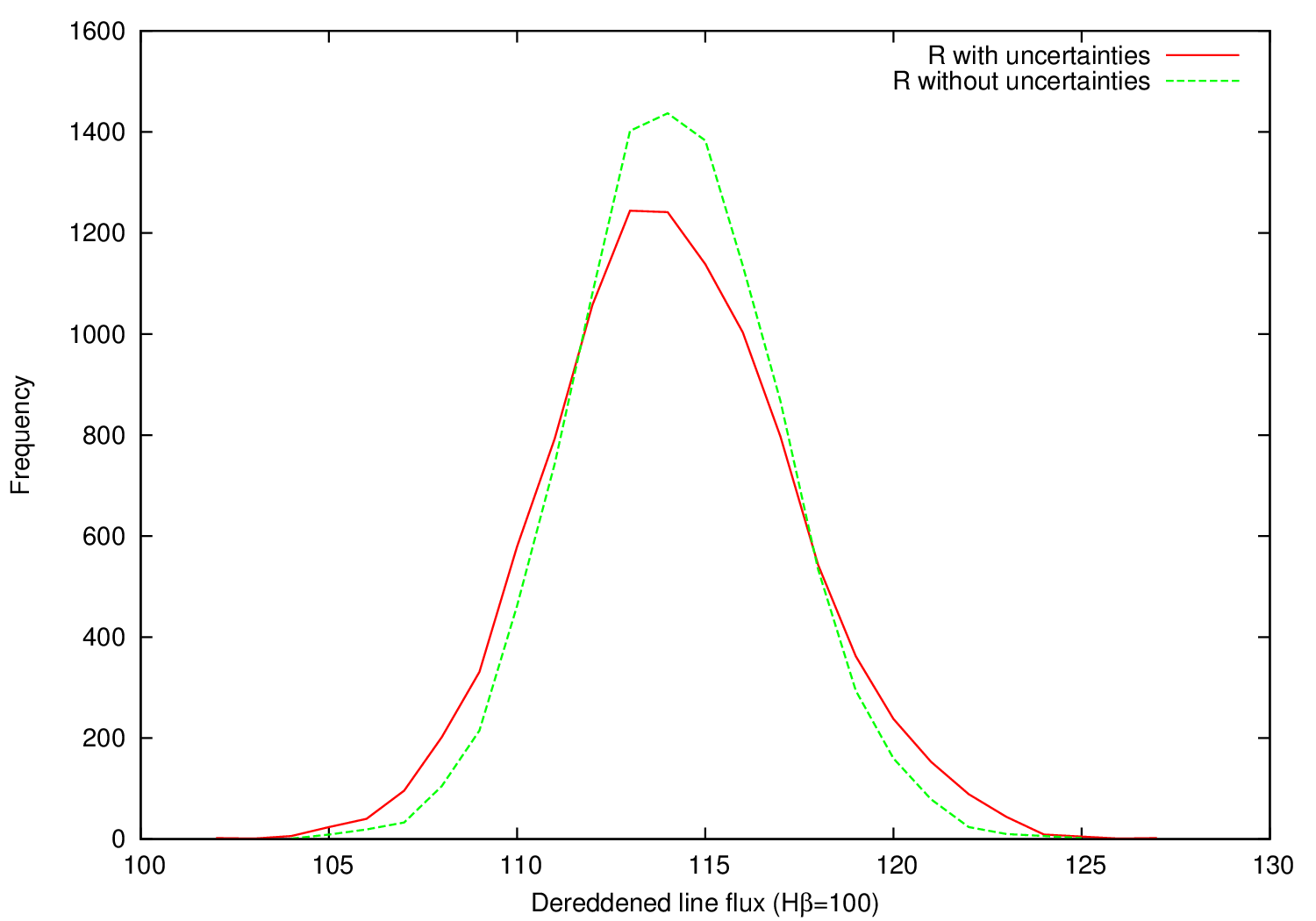}
\includegraphics[width=0.48\textwidth]{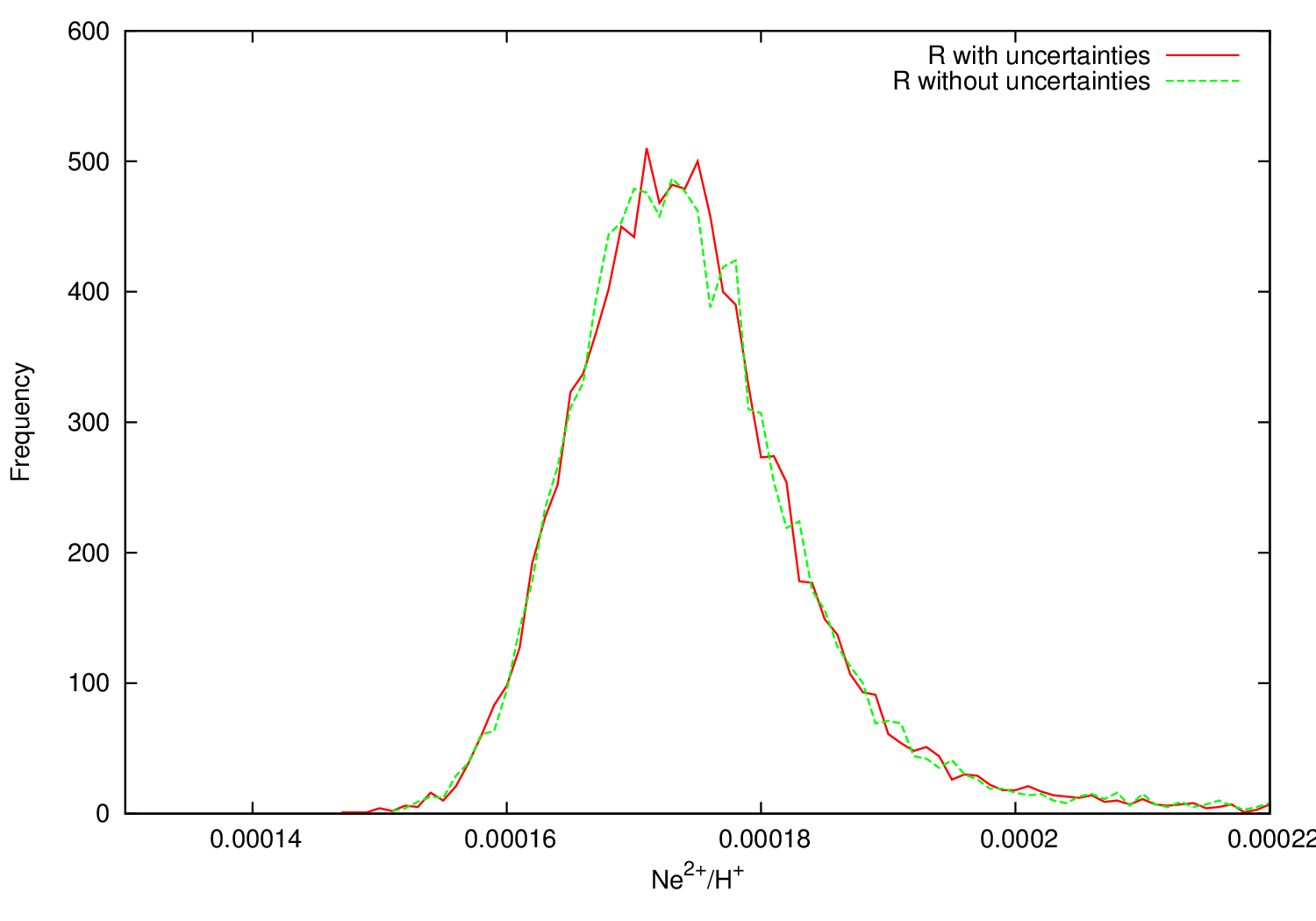}
\caption{The effect of an assumed uncertainty in R, the ratio of selective to total extinction, on the dereddened line flux of the [Ne~{\sc iii}] line at 3868{\AA} (l) and derived Ne$^{2+}$/H$^+$ abundance (r).} 
\label{R_effect}
\end{figure*}

We find that including the effect of an uncertainty in R has a noticeable effect on the probability distribution of dereddened line fluxes.  However, in the conversion from line fluxes into physical quantities, the statistical uncertainties arising from line flux measurements completely dominate, and the probability distributions are statistically identical whether R is assumed to be fixed or allowed to vary.  This result is shown in Figure~\ref{R_effect}, where we plot the probability distributions for the [Ne~{\sc iii}] 3868{\AA} dereddened line flux, and the abundance derived from it.  The effect on the derived abundance of the assumed uncertainty on R is negligible.

\section{Discussion}

Motivated by the necessity of properly understanding sources of uncertainties in analyses of photoionized nebulae, we have presented a new code for calculating chemical abundances in photoionized nebulae, which also robustly calculates the statistical uncertainties on the abundances determined.  The code is freely available and we welcome bug reports and feature requests.

Analytic methods of uncertainty propagation rely on assumptions that typically do not hold for real astronomical data, and therefore we have adopted a Monte Carlo approach.  We have shown that the analytic approach does not give a good estimate of the true statistical uncertainties.

We have provided in the code a means for accounting for the well known upward bias in flux measurements of weak lines, and we show that doing so results in reduced statistical as well as systematic uncertainties in abundance determinations.  This effect should not be ignored in any analyses of emission line spectra; in almost any data set, the weakest lines will be subject to the effect, and unless these lines are ignored in the analysis, then incorrect results will be obtained if the RP effect is not properly accounted for.  If, as is commonly the case, recombination lines make up the majority of the weak lines, then systematic misunderstandings may arise if the effect is neglected.  In the examples that we have analysed, however, this effect is too small to account for the magnitude of the discrepancy typically found between RL and CEL abundance determinations.

Finally we have shown that possible uncertainties in the value of R have a negligibly small effect on the uncertainties on the final quantities.

In the present analysis we have considered only statistical uncertainties.  The correct propagation of these, as we have seen, can reduce their final magnitude.  However, results obtained using {\sc neat} are of course subject to systematic uncertainties as well.  These have many potential origins: systematic uncertainties in atomic data and the choice of atomic data to use for each ion; the choice of interstellar extinction law; the choice of ionization correction scheme; the methodology of calculating the temperatures and densities to use for the abundance calculations; unjustified assumptions in the analysis such as assumptions of constant temperatures and densities; correction for underlying stellar absorption in certain types of nebula; and others that may yet be unknown.  As \citet{2011arXiv1109.2502K} point out, investigators starting from an identical line list may arrive at quite different results, depending on what choices they make regarding the sources of these systematic uncertainties.

Two questions which arise regarding these uncertainties are

\begin{enumerate}
\item Which systematic choices are the most important?
\item What is the statistical significance of the systematic uncertainties?
\end{enumerate}

Through a proper understanding of statistical uncertainties, it may be possible to answer these questions.  For example, by comparing the systematic uncertainty introduced by varying the choice of reddening law with the statistical uncertainty arising from line flux measurement, one could determine quantitatively whether or not the choice of reddening law is crucial or relatively inconsequential.  In a forthcoming paper we plan to extend this type of analysis to quantify the effects of systematic choices in terms of statistical uncertainties, and thus to be able to determine the relative importance of each of the systematic choices being made.

\section*{Acknowledgments}

We thank Professors Ian Howarth and Mike Barlow for useful discussions, Mahesh Mohan for testing an early version of the code, Bruce Duncan for helping us optimise the code, and the organisers of the workshop ``Uncertainties in atomic data and how they propagate in chemical abundances" for providing an extremely fruitful forum for interaction between atomic data providers and users.  We also thank the anonymous referee for their very useful comments and suggestions.  This work was co-funded under the Marie Curie Actions of the European Commission (FP7-COFUND). DJS is supported by an NSERC Discovery Accelerator Grant.  PS is supported by the ESO Studentship Programme.

\bibliographystyle{mn2e}
\bibliography{NEAT_paper}

\label{lastpage}

\end{document}